\documentclass[aps,prb,preprint]{revtex4}
\usepackage[dvipdfm]{graphicx} % Include figure files
\usepackage{bm}% bold math
\usepackage{booktabs}% tabs
\usepackage{amsmath,amsthm,amssymb}
\usepackage{ascmac}

\begin{document}

\title{Spin-charge transport driven by magnetization dynamics on disordered surface of doped topological insulators}
\author{K. Taguchi, K. Shintani, and Y. Tanaka}

\affiliation{Department of Applied Physics, Nagoya University, Nagoya, 464-8603, Japan
\\
and CREST, Japan Science and Technology Corporation (JST), Nagoya 464-8603, Japan} %\\ $^2$Department of Physics, Tokyo Institute of Technology, Tokyo, 152-8551, Japan}
%\email{taguchi@rover.nuap.nagoya-u.ac.jp} %taguchi-katsuhisa@ed.tmu.ac.jp
\date{\today}
%
% abstract 
\begin {abstract}
We theoretically study the spin and charge generation along with the electron transport on a disordered surface of a doped three-dimensional topological insulator/magnetic insulator junction by using Green's function techniques. 
We find that the spin and charge current are induced by not only local but also by nonlocal magnetization dynamics through nonmagnetic impurity scattering on the disordered surface of the doped topological insulator. 
We also clarify that the spin current as well as charge density are induced by spatially inhomogeneous magnetization dynamics, and the spin current diffusively propagates on the disordered surface. 
Using these results, we discuss both local and nonlocal spin torques before and after the spin and spin current generation on the surface, and provide a procedure to detect the spin current.
%\\ \\ 
%PACS numbers: MR
%       72.25.-b, %Spin polarized transport (for spin polarized transport devices, see 85.75.-d) 
%       73.43.Qt, %Magnetoresistance 
%       75.47.-m, %Magnetotransport phenomena
%       75.76.+j        Spin transport effects (for devices exploiting spin polarized transport, see 85.75.Hh, 85.75.Mm, and 85.75.Ss)
%------------- DCSE
%PACS numbers: 
%       85.75.-d %Magnetmelectronics;spintronics:device exploiting spin polarized transport or integrated magnetic fields 72.25.-b, %Spin polarized transport (for spin polarized transport devices, see 85.75.-d) 
%       75.47.-m, %Magnetotransport phenomena
%       72.25.-b, %spin polarized transport (for spin polarized transport devices, see 78.75.-d) 
%       [ 75.76.+j      Spin transport effects (for devices exploiting spin polarized transport, see 85.75.Hh, 85.75.Mm, and 85.75.Ss)]
%\begin{description}
%\item[] \hspace{8.cm}  Subject Areas: Condensed Matter Physics, 
%	\\ \hspace{9.8cm} Spintronics,
%	\\ \hspace{9.8cm} Topological  insulators
%PACS numbers: 85.75.-d, 75.47.-m, 72.25.-b %\verb+\pacs{#1}+ command.
%\item[]Keywords: %Ultrafast magnetic vortex switching, Magnetooptical effect, inverse Faraday effect, Spin Berry phase   
%\end{description}
\end{abstract}
%\pacs{}
%\keywords{}

%\keywords{}
\maketitle

\section{Introduction}
% Introduction to generation spin current and f due to magnetization dynamics
In spintronics, the mutual control of the direction and the flow of spin is 
a central issue for wide applications.
The flow of spin, i.e., spin current, 
is the difference between the currents of up and down-spin 
conduction electrons.
It is known that 
the spin current is induced in the setup of the ferromagnetic metal (FM)/normal metal (NM) junction\cite{rf:Tserkovnyak02,rf:ohe07,rf:saitoh06}.
Its origin is due to the magnetization dynamics in ferromagnet, 
which transfers the spin angular momentum of the magnetization 
into that of the conduction electrons.
The transfer of the spin angular momentum is called spin-pumping.
Here, the spin-pumping of magnetization dynamics generates the spin-current 
in the NM, and the spin-current can be converted into charge current through 
spin-orbit interactions\cite{rf:Tserkovnyak02,rf:ohe07,rf:saitoh06,rf:Stern92,rf:Takeuchi08}.

% Introduction to TI  
Topological insulator (TI) is a new class of materials 
which has a gapless surface state, dubbed as the helical surface state, in which the spin and momentum  are locked by 
the spin-orbit interactions\cite{rf:Hasan10,rf:Qi11,rf:Ando13}. 
On the surface of the TI, 
the direction of charge current and that of the spin of 
conduction electrons can be 
mutually manipulated by an applied electromagnetic field 
through the spin-momentum locking. 
There have been many theoretical works 
in hybrid systems including superconducting 
junctions on the surface of TI 
stimulated  by the exotic surface state\cite{rf:Qi11,rf:Qi08,rf:FuKane,rf:Akhmerov,rf:Tanaka}.
In the TI/FM junctions, 
the anomalous charge-spin 
transport \cite{rf:Qi08,rf:Nomura10,rf:Garate10,rf:Ueda12}, 
the anomalous tunnel conductance
\cite{rf:Yokoyama10R,rf:Mondal10,rf:Kong11,rf:Ma12,rf:Culcer12}, the giant 
magneto resistance\cite{rf:Burkov10,rf:Schwab11,rf:Taguchi14}, and the 
current-induced spin-transfer torque\cite{rf:Yokoyama10,rf:sakai14} have been 
studied up to now.
The exotic phenomena are triggered in the presence of static magnetization and 
an applied electromagnetic field. 
The magnetization of the ferromagnet plays the role of an effective vector 
potential for conduction electrons, which is like a vector potential of 
electromagnetic fields.
Owing to the effective vector potential,  
the time-derivative of the magnetization can be regarded as an effective electric field, and the magnetization dynamics generates charge current on the surface 
of the TI/FM junction even in the absence of electromagnetic fields\cite{rf:Mellnik,rf:Deorani,rf:Shiomi14,rf:Jamali}.
This is called as the spin-charge conversion.
The direction of the induced charge current is perfectly 
perpendicular to the magnetization dynamics 
due to the spin-momentum locking. 
The relation between the direction of the magnetization and the induced charge 
current can be a characteristic property on the surface of the TI. 
The property of the spin-pumping on the surface of TI can be applicable for spintronics devices.  

Existing works of the spin-charge conversion have been done 
in the case of a clean surface of the TI, namely the ballistic transport regime.
However, the actual charge transport on the surface of the TI 
is in the diffusive regime due to the 
nonmagnetic impurity scattering\cite{rf:Burkov10,rf:Schwab11,rf:Taguchi14,rf:Yokoyama10,rf:sakai14,rf:Liu,rf:Fischer,rf:Yokoyama}. 
Since Burkov et al., have predicted not only the local 
but also the nonlocal current on the disordered surface of the TI in the presence of the applied electric field\cite{rf:Burkov10}; 
we can naturally expect nonlocal 
current is driven by the magnetization dynamics 
even on the disordered surface of the TI in the presence of the magnetization dynamics.

% To do in this paper
In this article, we study the charge-spin transport due to the magnetization dynamics on the disordered surface of the three-dimensional doped TI/magnetic insulator (MI) junction, as shown in Fig. 1, 
where we show that charge current and spin polarization on the surface of the TI are induced not only by a local, but also by a nonlocal magnetization dynamics.
Besides this, we clarify that the spin current is driven by the dynamics of the spatially inhomogenous magnetization, and 
the spin current diffusively propagates on the surface. 
The magnitude of the spin current reflects the spatially inhomogenous spin structure of the MI.
The directions of the spin flow and the spin projection of the spin current 
are perfectly linked by the spin-momentum locking on the surface of TIs. 
The present features may serve as a guide to fabricate future spintronics 
devices based on the surface of TIs with magnetic substance.

The merit of the choice of MI on the surface of the TI instead of 
metallic ferromagnet is 
to prevent the induced charge current going through 
the bulk of the MI. 
Then, we can focus on the charge transport on the 
surface of TI. 
Besides, the Gilbert damping constant in MIs tends to be smaller than that in ferromagnetic metals. 
The small value of the damping in MIs can be useful for the detection of the spin current on the disordered surface of the TI, as discussed in sec. \ref{sec:5-c}
%The reason is that the Gilbert damping constant (or the half-width value of the magnetic permeability) in MIs is smaller (or shorter) than that in ferromagnetic metals.

\section{Model}
We consider conduction electrons coupled to an effective localized spin on the disordered surface of the three-dimensional doped TI attached with the MI, 
as shown in Fig. 1.
The setup in Fig. 1 is similar to a system, where conduction electrons 
couple with the magnetic moments of ferromagnetic metals deposited on the surface of the TI\cite{rf:Qi08,rf:Yokoyama10R,rf:Mondal10}. 
We expect that on the surface of the TI, the effective localized spin ($\bm{S}$) can be produced from the magnetization in the MI through magnetic proximity effects. 
%Then, we assume that properties of conduction electrons on the surface of the T%I is not dramatically changed by attaching the MI.
%The localized spin we consider is produced on the surface of the TI by the proximity effect from the MI. 
In the following, we use 
the Hamiltonian, describing the surface of the TI with MI, given by 
\begin{align} \label{eq:2-0-1} %\ref{eq:2-0-1}
\mathcal{H} = \mathcal{H}_{\rm{TI}} + \mathcal{H}_{sd}  + V_{\textrm{imp}},
\end{align}
where the first term in Eq. (\ref{eq:2-0-1}),  $\mathcal{H}_{\rm{TI}}$ is Hamiltonian of the conduction electrons on one of the surface of the doped TI without $\bm{S}$ as
\begin{align} \label{eq:2-0-2} %\ref{eq:2-0-2}
\mathcal{H}_{\rm{TI}} & =   \int d\bm{x} \psi^\dagger [ -i\hbar v_{\rm{F}} (\hat{ \bm{\sigma}} \times \bm{\nabla})_z - \epsilon_{\textrm{F}} ]\psi, 
\end{align}
Here, $\psi^\dagger \equiv  \psi^\dagger(\bm{x},t) = (\psi^\dagger_{\uparrow} \ \psi^\dagger_{\downarrow})$, and $\psi$ are the creation and annihilation operators of the conduction electron, respectively (where indices $\uparrow$ and $\downarrow$ represent spin), $\epsilon_{\rm{F}}$ is the Fermi energy, and $v_{\rm{F}}$ is the Fermi velocity of the bare electron on the surface of the doped TI. 
The $\hat{\bm{\sigma}}$ is the Pauli matrices in spin space. 
%
% Explain the second term 
The second term of Eq. (\ref{eq:2-0-1}), $\mathcal{H}_{sd}$, 
shows the exchange interaction between the conduction electron spin $\bm{s} = \frac{1}{2} \psi^\dagger \bm{\sigma}\psi$ and the localized spin $\bm{S}$ on the disordered surface of the doped TI, as described by 
\begin{align} 
	\label{eq:2-0-3} %\ref{eq:2-0-3}
\mathcal{H}_{sd} & = - \int d\bm{x} J_{sd}  \psi^\dagger  \bm{S}\cdot \hat{\bm{\sigma}}  \psi,
%	\\
%	
%\mathcal{H}_{sd}^z & = - \int d\bm{x} J_{sd}  \psi^\dagger   S^z \sigma^z  \psi,
\end{align}
where $J_{sd} >0$ is the exchange coupling constant.
The localized spin $\bm{S}$ can be described by the magnetization of the MI as $\bm{S}=-(S/M)\bm{M}$, 
where $S$ and $M$ are the magnitude of the localized spin and of the magnetization, respectively.
We consider that in general, the localized spin $\bm{S} \equiv \bm{S}(\bm{x},t)$ depends on the time and position on the surface of the TI.
The $\bm{S}(\bm{x},t)$ changes slowly compared with the electron transport relaxation time ($\tau$) and varies in space compared with the electron mean-free path ($\ell$).
%
% expectation
We expect that from the Eqs. (\ref{eq:2-0-2})-(\ref{eq:2-0-3}), the in-plane component of the localized spin, $\bm{S}^\parallel \equiv \bm{S} - S^z \bm{z}$, can play the role of the effective vector potential for the conduction electrons on the surface. 
The out-of plane component of the localized spin $S^z$ plays a role to open 
the energy gap of the dispersion on the surface of the doped TI. 
We assume that the band gap opened by $S^z$ is smaller than the Fermi energy on the surface of the doped TI, $i.e., \epsilon_\textrm{F} - J_{sd} S^z>0$.
The third term of Eq. (\ref{eq:2-0-1}), 
\begin{align} \label{eq:2-0-4} %\ref{eq:2-0-4}
V_{\textrm{imp}} =  \sum^{N_i}_{j=1}  \int d\bm{x} U_{\textrm{i}} \psi^\dagger \psi,
\end{align}
represents nonmagnetic impurity scattering on the disordered surface of the doped TI.
The impurity scattering causes the relaxation time $\tau$ of the transport of conduction electrons on the surface of the TI.
Here $U_{\textrm{i}}=u_{\textrm{i}} \delta (\bm{x} - \bm{r}_j)$ is a delta-function type potential, $u_{\textrm{i}}$ is a potential energy density, $\bm{r}_j$ is the position of impurities, and $N_\textrm{i}$ shows the number 
of impurities. 
The contribution of $V_{\rm{i}}$ can be treated by the impurity average.

% Condition we calculate
We will calculate the spin current and charge current due to the spin-pumping in the linear response to $\bm{S}$ under the condition $J_{sd}  \ll \hbar/\tau$. 
%We will calculate the spin current and charge current due spin-pumping in the linear response to the localized spin, as shown in Sec. III and IV, respectively. 
We expect that the condition could be realized on a metallic disordered surface of the TI, which satisfies $\hbar/(\epsilon_\textrm{F}\tau) \ll 1$.
For example, 
the exchange coupling $J_{sd}$ can be estimated by $J_{sd} \simeq 6$meV \cite{rf:Shiomi14} for  Ni$_{81}$Fe$_{19}$/Bi$_2$Te$_3$ junction. 
If $\frac{\hbar }{\epsilon_\textrm{F}\tau} < 10^{-2}$ is satisfied, 
the perturbation  can be accessible. 
\begin{figure} \label{fig:1} % \ref{fig:1}
\includegraphics[scale=.5]{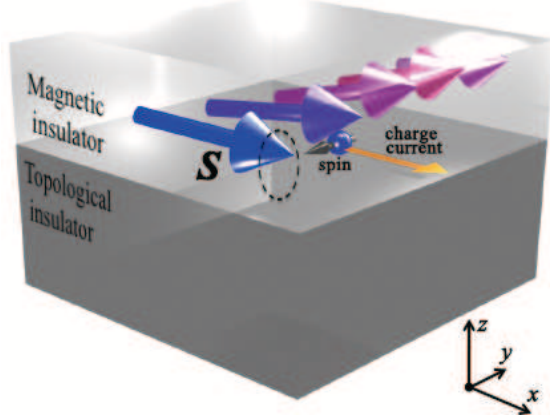}\centering
\caption{(Color Online) A setup of spin-charge current generation due to magnetization dynamics in MIs deposited on disordered surface of a doped TI. }
\label{f1}
\end{figure}

\subsection{Renormalization of the Fermi velocity}%Green's function on disordered surface of doped topological insulator}
Green's function on disordered surface of doped TI can be described by using $\mathcal{H}_{\textrm{TI}}$ including $V_{\textrm{imp}}$ within the 
self-consistent Born approximation of $V_{\textrm{imp}}$ as
\begin{align}
\label{eq:2-1-5} %\ref{eq:2-1-5}
\hat{g}_{\bm{k},\omega}
	& = [\hbar \omega - \{\hbar v_\textrm{F} (\hat{\bm{\sigma}} \times \bm{k})_z - \epsilon_{\textrm{F}} \} -\hat{\Sigma}_{\bm{k},\omega}  ]^{-1},
\end{align}
where $\hat{\Sigma}_{\bm{k},\omega}$ is the self-energy within the Born approximation given by 
\begin{align}
\label{eq:2-1-6} %\ref{eq:2-1-6}
\hat{\Sigma}_{\bm{k},\omega}
	& = n_i \sum_{\bm{k'}} |u_{\bm{k}-\bm{k'}}|^2  \hat{g}_{\bm{k'},\omega}.
\end{align}
Here, $\hat{\Sigma}_{\bm{k},\omega}$ satisfies the Ward-Takahashi identity\cite{rf:book2}.
%:
%$\hbar \hat{v}_i + \partial_{k_i} \hat{\Sigma}_{\bm{k},\omega} = \hat{\Lambda}_{i, {\bm{k},\omega}}$, where  $\hat{\Lambda}_i $ is the vertex function of the ladder diagram of $V_{\textrm{imp}}$ and $\hat{v}_i=  v_{\textrm{F}} (\bm{z}\times \hat{\bm{\sigma}})$ is the velocity operator on the surface of the TI.
To estimate the value of $\hat{\Sigma}_{\bm{k},\omega}$, we consider the $\bm{k'}$-dependence of $u_{\bm{k}-\bm{k'}}$, which plays the role to prevent the ultraviolet-divergence over a large momentum in the $\bm{k'}$-integration\cite{rf:Fujimoto13, rf:sakai14}. 
When $\hat{\Sigma}_{\bm{k},\omega}$ can be described by 
$2 \times 2$ matrix  
$\hat{\Sigma}_{\bm{k},\omega} = \Sigma_0 + \hat{\Sigma}^\parallel + \Sigma^z \hat{\sigma}^z$, $\hat{\Sigma}$ is estimated, where $\Sigma_0$ and $\Sigma^z$ are independent of $\bm{k}$, while  
$\Sigma^\parallel \equiv \Sigma^x \hat{\sigma}^x + \Sigma^y \hat{\sigma}^y$ depends on $\bm{k}$.
Then, the Green's function $\hat{g}_{\bm{k},\omega}$ is described by $\Sigma$ within the Born approximation as\cite{rf:Fujimoto13, rf:sakai14}
\begin{align} \label{eq:2-1-7} %\ref{eq:2-1-7}
\hat{g}^r_{\bm{k},\omega}
	& = \biggl[ \hbar \omega - \{\hbar \tilde{v}_\textrm{F} (\hat{\bm{\sigma}} \times \bm{k})_z - \epsilon_{\textrm{F}} \} + \frac{i\hbar}{2\tau} \biggr]^{-1},
\end{align}
where $\hat{g}^r_{\bm{k},\omega}$ represents the retarded Green's function.
From Eq. (\ref{eq:2-1-7}), the Fermi velocity is renormalized by $\Sigma^\parallel$, and 
the renormalized Fermi velocity is represented by $\tilde{v}_\textrm{F}=v_\textrm{F}/(1+\xi)$, where $\xi=n_i u_0^2/(4\pi \hbar^2 v_\textrm{F}^2)$ is a small value depending on the relaxation time\cite{rf:estimation}. 
The last term in Eq. (\ref{eq:2-1-7}) is caused by the retarded component of $\textrm{Im}[\Sigma_0]$.

Equation (\ref{eq:2-1-7}) indicates the Green's function on the disordered surface of the doped TI estimated within the Born approximation of $V_{\textrm{imp}}$.
Therefore, we could expect that $\hbar \tilde{v}_\textrm{F} (\hat{\bm{\sigma}} \times \bm{k})_z - \epsilon_{\textrm{F}}$ in the Eq. (\ref{eq:2-1-7}) corresponds to the dispersion on the disordered surface of the TI.
The dispersion is different from that of $\mathcal{H}_{\textrm{TI}}$ without $V_{\textrm{imp}}$. 
In the following work, 
we will use an effective Hamiltonian $\tilde{\mathcal{H}}_{\textrm{TI}}$ 
obtained by replacing $v_{\rm{F}}$ with $\tilde{v}_{\rm{F}} $ in Eq. (\ref{eq:2-0-2}).
This replacement is needed for satisfying the charge conservation law on the disordered surface of the doped TI\cite{rf:discussion1}.  
%\begin{align}
%\mathcal{H}_{\textrm{TI}} \to \tilde{\mathcal{H}}_{\textrm{TI}}
%	& =  \int d\bm{x} \psi^\dagger [ -i\hbar \tilde{v}_{\rm{F}} (\hat{ \bm{\sigma}} \times \bm{\nabla})_z - \epsilon_{\textrm{F}} ]\psi 
%\end{align}

%\bibitem{rf:discussion1}
%If we do not take the replacement $v_{\rm{F}} \to \tilde{v}_{\rm{F}} $ in Eq. (\ref{eq:1-2}), the velocity operator on the surface of the TI becomes $\hat{v} = v_{\textrm{F}} (\bm{z}\times\hat{\bm{\sigma}})$ and the charge current is shown in Eqs. (\ref{eq:7-44})-(\ref{eq:8-45}) multiple of $v_\textrm{F}/\tilde{v}_{\rm{F}}$. 
%Then, the charge current does not satisfy the charge conservation law.

%Spin current due to localized spin dynamics
\section{Spin current due to magnetization dynamics}
In this section, 
we show spin current driven by magnetization dynamics on the disordered surface of the doped TI/MI junction. 
Here, the spin current and charge density are mutually related each other, because of the spin-momentum locking on the surface of the TI.
 
\subsection{Definition of spin current on the surface of topological insulators}
In order to derive the spin current on the disordered surface of the doped TI, 
we demonstrate the definition of the spin current. 
The spin current $j_{i}^\alpha$ is defined from  
\begin{align}  \label{eq:3-1-8} %\ref{eq:3-1-8}
\partial_t s^\alpha & + \nabla_i j_{i}^\alpha  =  \mathcal{T}^\alpha,
\end{align}
where $s^\alpha =\frac{1}{2} \langle \psi^\dagger \hat{\sigma}^\alpha \psi \rangle$ is the spin density, 
$j_{i}^\alpha$ shows the spin current density, 
and $\mathcal{T}^\alpha$ is the spin relaxation torque on the surface.
Here the subscript and superscript of $j_{i}^\alpha$ represent the direction of flow and spin of the spin current, respectively.  
From Eqs. (\ref{eq:2-0-1})-(\ref{eq:2-1-5}), $j_{i}^\alpha$ and $\mathcal{T}^\alpha$ are given by 
\begin{align}
	\label{eq:3-1-9} %\ref{eq:3-1-9}
j_{i}^\alpha  = & \frac{ \tilde{v}_\textrm{F} }{2} \epsilon_{z \alpha i} \langle \psi^\dagger  \psi \rangle
	 =  \frac{ \tilde{v}_\textrm{F}  }{2e} \epsilon_{z  \alpha i} \rho_e.
\end{align}
with the Levi-Civita symbol $\epsilon_{z  \alpha i}$. 
In the above equation, we used the commutation relation: 
\begin{align}
\notag
\begin{matrix}
[\psi^\dagger, \mathcal{H}] 
	 = -i\hbar \tilde{v}_\textrm{F}  \epsilon_{z\alpha \ell }(\nabla_\ell \psi^\dagger) \sigma^\alpha
		+ J_{sd} \psi^\dagger S^\alpha \sigma^\alpha , 
	\\ 
	\notag 
[\psi, \mathcal{H}]  = -i\hbar \tilde{v}_\textrm{F}  \epsilon_{z\alpha \ell } \sigma^\alpha (\nabla_\ell \psi)
	- J_{sd}  S^\alpha \sigma^\alpha \psi.
\end{matrix}
\end{align}
From Eq. (\ref{eq:3-1-9}), the spin current is proportional to the charge density $\rho_e = e \langle \psi^\dagger \psi \rangle$, where $e<0$ is the charge of electrons.
Moreover, the directions of the spin and flow of the spin current are perpendicular to each other because of the spin-momentum locking.
The spin relaxation torque is derived from Eqs. (\ref{eq:3-1-8})-(\ref{eq:3-1-9}). 
The torque can be separated as 
\begin{align}
	\label{eq:3-1-10}%\ref{eq:3-1-10}
\mathcal{T}^\alpha 
	= & \mathcal{T}^\alpha_{\textrm{TI}}  + \mathcal{T}^\alpha_{\textrm{sd}}, 
\end{align}
where 
$\mathcal{T}^\alpha_{\textrm{TI}} $ and $\mathcal{T}^\alpha_{sd}$ are spin relaxation torque caused by $\mathcal{H}_{\textrm{TI}}$ and $\mathcal{H}_{sd}$, respectively.
Here, $\mathcal{T}^\alpha_{\textrm{TI}} $ and $\mathcal{T}^\alpha_{sd}$ are given by
\begin{align}
	\label{eq:3-1-11}%\ref{eq:3-1-11}
\mathcal{T}^\alpha_{\textrm{TI}}
	= & \frac{i \tilde{v}_\textrm{F}  }{2} \epsilon_{\beta z \ell} \epsilon_{\beta \alpha \nu} 
		\langle (\nabla_\ell \psi^\dagger) \hat{\sigma}^\nu \psi -  \psi^\dagger \hat{\sigma}^\nu \nabla_\ell \psi \rangle,
	\\ \label{eq:3-1-12}%\ref{eq:3-1-12}
\mathcal{T}^\alpha_{\textrm{sd}}
	= &  \frac{2J_{sd}}{\hbar} \epsilon_{ \nu \beta \alpha}  s^\nu S^\beta.
\end{align} 
We note that  the definition of spin current depends on that of the spin relaxation torque\cite{rf:Culoer,rf:Shi,rf:Nakabayashi}.
For example, we consider the case when the spin relaxation torque can be described by $\tau^{\alpha} = \mathcal{T}^{\alpha} + \nabla_i \mathcal{P}^{\alpha}_{i}$, 
where the polarization $\mathcal{P}^{\alpha}_{i}$ is an arbitrary vector 
with $\nabla_i \mathcal{P}^{\alpha}_{i}=0$, 
whose index $i$ and $\alpha$ represent the direction of the polarization in the real space and that of the spin in the spin space, respectively. 
Then, the spin current $J_{i}^\alpha$ can be also represented by 
$\mathcal{J}_{i}^\alpha = j_{i}^\alpha +  \mathcal{P}^{\alpha}_{i} $ and $\mathcal{J}_{i}^\alpha$ satisfies the conservation law as 
$\partial_t s^\alpha + \nabla_i \mathcal{J}_{i}^\alpha = \tau^\alpha$.
We discuss the spin current defined in Eq. (\ref{eq:3-1-9}).
To consider the spin current and the spin relaxation torque, 
we calculate the charge density and spin density in the following subsections.

%%%%%%%%%%%%%%%%%%%
\subsection{Charge density}% $\rho_e = e \langle \psi^\dagger \psi\rangle $
First, we will calculate the charge density $\rho_e$ in the linear response to $\bm{S}$.
$\rho_e$ is described by using the lesser component of the Keldysh-Green's function, $-i\hbar G^{<} (\bm{x},t, \bm{x},t) = \langle \psi^\dagger(\bm{x},t) \psi(\bm{x},t) \rangle$ in the same position and time as 
\begin{align} \label{eq:3-2-13} %\ref{eq:3-2-13}
\rho_e & =  -i\hbar e \ {\textrm{tr}} \bigl[ \hat{G}^{<} (\bm{x},t, \bm{x},t)\bigr].
\end{align} 
Hence, $\rho_e$ is given by   
\begin{align} 
	\label{eq:3-2-14} %\ref{eq:3-2-14}
\rho_e	& =  \frac{i \hbar e  J_{sd}}{L^2} \sum_{\bm{q},\Omega}e^{i (\Omega t-\bm{q}\cdot\bm{x} )} 
		{\rm{tr}}[ \hat{\Pi}_{0 \nu} (\bm{q},\Omega) S^{\nu}_{\bm{q},\Omega} ], 
\end{align}
where $L^2$ is the area of the disordered surface of the TI, and $\bm{q}=(q_x, q_y)$ and $\Omega$ indicate the momentum and frequency of the localized spin $S_{\bm{q},\Omega}^{\nu}$ ($\nu=x,y,z$), respectively.
Here, 
the charge-spin correlation function $\hat{\Pi}_{0 \nu} $ is given by  
\begin{align}
	\label{eq:3-2-15} %\ref{eq:3-2-15}
\hat{\Pi}_{0 \nu}(\bm{q},\Omega)  
	& = \sum_{\bm{k}, \omega} [ \hat{g}_{ \bm{k}-\frac{\bm{q}}{2},\omega-\frac{\Omega}{2}} \hat{\Lambda}_\nu (\bm{q},\Omega) \hat{g}_{ \bm{k}+\frac{\bm{q}}{2},\omega+\frac{\Omega}{2}} ]^<,
\end{align}
where $\hat{g}_{ \bm{k}\pm\frac{\bm{q}}{2},\omega\pm\frac{\Omega}{2}}$ is the non-perturbative Green's function of $\mathcal{H}_{\textrm{TI}}$ including $V_{\rm{i}}$, which is taken into account within the Born approximation.
The retarded (advanced) Green's function $\hat{g}^r_{\bm{k},\omega}$ ($\hat{g}^a_{\bm{k},\omega} =  [\hat{g}^r_{\bm{k},\omega}]^\dagger)$ is given by 
\begin{align*}
\hat{g}^r_{\bm{k},\omega} 
	 = \left[ \hbar \omega +\epsilon_{\rm{F}} -\hbar \tilde{v}_\textrm{F}  \hat{ \bm{\sigma}} \cdot (  \bm{k} \times \bm{z})  + i \hbar /(2\tau) \right]^{-1},
\end{align*}
where $\hbar/(2\tau)=\pi n_{\rm{i}} u_{\rm{i}}^2 \nu_e/2 $ represents the self-energy due to $V_{\textrm{i}}$ within the Born approximation.
The vector 
$\hat{\Lambda}_\nu$ in Eq. (\ref{eq:3-2-15}) is the vertex function, which is described by %the $2 \times 2$ Pauli matrix as 
\begin{align}
	\label{eq:3-2-16} %\ref{eq:3-2-16}
	\hat{\Lambda}_\gamma (\bm{q},\Omega)
		& =  \hat{\sigma}_\gamma 
			+ \sum_{\zeta=0,x,y,z}[ \tilde{\Gamma}(\bm{q},\Omega) + [\tilde{\Gamma}(\bm{q},\Omega)]^2 + \cdots ]_{\gamma \zeta} \hat{\sigma}_\zeta.
\end{align}
Here $\hat{\sigma}_0 = \hat{1}_{2\times2}$ is the identity matrix and 
$\tilde{\Gamma}_{\gamma \zeta}$ is given by $\hat{ \Gamma }_\gamma$:
\begin{align} \notag 
	\hat{ \Gamma }_\gamma(\bm{q},\Omega)
		&\equiv n_{\rm{i}} u_{\rm{i}}^2  \sum_{\bm{k}}  \hat{g}_{\bm{k}-\frac{\bm{q}}{2},\omega-\frac{\Omega}{2}} \hat{\sigma}_\gamma \hat{g}_{\bm{k}+\frac{\bm{q}}{2},\omega+\frac{\Omega}{2}}
		\\ \label{eq:3-2-17} %\ref{eq:3-2-17}
		& = \tilde{\Gamma}_{\gamma \zeta} \hat{\sigma}_\zeta.
\end{align}
%%%
The correlation function $\hat{\Pi}_{0 \nu} $ can be decomposed into the retarded and advanced Green's function by using the formula $ \hat{g}^<_{\bm{k},\omega} = f_\omega (\hat{g}^a_{\bm{k},\omega} - \hat{g}^r_{\bm{k},\omega} )$ \cite{rf:book1}, where $f_\omega$ is the Fermi distribution function. 
Using the formula, we can estimate the correlation function $\hat{\Pi}_{0 \nu} $ on the surface of the doped TI, $i.e., \hbar /(\epsilon_\textrm{F} \tau ) \ll 1$ regime as $\hat{\Pi}_{0 \gamma} = \hat{\Pi}^{\rm{ra}}_{0 \gamma} + o(\hbar /(\epsilon_{\textrm{F}}\tau ))$, where $\hat{\Pi}^{\rm{ra}}_{0 \gamma}$ is represented by 
\begin{align}
	\label{eq:3-2-18} %\ref{eq:3-2-18}
\hat{\Pi}^{\rm{ra}}_{0\gamma} 
	& = \sum_{\bm{k}, \omega} (f_{\omega+\frac{\Omega}{2}} -f_{\omega-\frac{\Omega}{2}} )   \hat{g}^r_{\bm{k}-\frac{\bm{q}}{2},\omega-\frac{\Omega}{2}} \hat{\Lambda}^{\rm{ra}}_\gamma  \hat{g}^a_{\bm{k}+\frac{\bm{q}}{2},\omega+\frac{\Omega}{2}}.
\end{align}
%%%%
Here $\hat{\Lambda}_\gamma^{\rm{ra}}$ is given by the Pauli matrix as 
\begin{align} 
	\label{eq:3-2-19} %\ref{eq:3-2-19}
\hat{\Lambda}_{\gamma}^{\rm{ra}}
		& = \sum_{\zeta=0,x,y,z} \bigl[ \hat{1} + \tilde{\Gamma}^{\rm{ra}} + (\tilde{\Gamma}^{\rm{ra}})^2  + \cdots \bigr]_{\gamma \zeta} \hat{\sigma}_\zeta
		 \equiv \sum_{\zeta} \tilde{\Lambda}^{\rm{ra}}_{\gamma \zeta } \hat{\sigma}_\zeta
		,  
		\\ 
		\label{eq:3-2-20} %\ref{eq:3-2-20}
\hat{\Gamma}_{\gamma}^{\rm{ra}}
	& =  n_{\rm{i}} u_{\rm{i}}^2  \sum_{\bm{k}} \hat{g}^{r}_{\bm{k}-\frac{\bm{q}}{2},\omega-\frac{\Omega}{2}} \hat{\sigma}_\gamma \hat{g}^{a}_{\bm{k}+\frac{\bm{q}}{2},\omega+\frac{\Omega}{2}}.
\end{align}
%We note that $\hat{\Lambda}_{\gamma}^{\rm{ra}}$ and $\hat{\Gamma}_{\nu}^{\rm{ra}}$ ($\nu=0,x,y,z$) are $2\times 2$ matrix described by the $2\times 2$ Paul matrix. 
%
Using Eqs. (\ref{eq:3-2-19})-(\ref{eq:3-2-20}) under the 
condition $\Omega\tau\ll1$,  
we can calculate $\hat{\Pi}^{\rm{ra}}_{0\gamma}$ in the low-temperature limit.  
%Next, we expand $f_\omega$ with the condition $\Omega\tau\ll1$.
%
Besides this, we can calculate the response function 
$\hat{\Pi}^{\textrm{ra}}_{0\gamma} $ by postuating 
$\Omega \tau \ll1$, $q\ell \ll1$, 
and $q_x^{2}=q_y^{2}=q^{2}/2$.  
$\hat{\Pi}^{\textrm{ra}}_{0\gamma} $ is given by 
\begin{align} 	\label{eq:3-2-21} %\ref{eq:3-2-21}
 \hat{\Pi}^{\rm{ra}}_{0\gamma} 
	= &\frac{-\Omega}{2\pi} \sum_{\zeta=0,x,y,z} \hat{\Gamma}_\zeta^{\textrm{ra}}  \tilde{\Lambda}^{\textrm{ra}}_{\gamma \zeta},
\end{align}
where $\hat{\Gamma}_\zeta^{\textrm{ra}} \equiv \sum_{\nu=0,x,y,z} \tilde{\Gamma}_{\zeta\nu}^{\textrm{ra}} \sigma_\nu$ can be expressed by $4\times 4$ matrix $\tilde{\Gamma}^{\textrm{ra}}$ as 
\begin{align}  
%	\label{eq:3-2-22} %\ref{eq:3-2-22}
%\begin{pmatrix} \hat{\Gamma}^{\textrm{ra}}_0 \\ \hat{\Gamma}^{\textrm{ra}}_x \\ \hat{\Gamma}^{\textrm{ra}}_y \\ \hat{\Gamma}^{\textrm{ra}}_z \end{pmatrix}
%	& = \tilde{\Gamma}^{\textrm{ra}}
%	\begin{pmatrix} \hat{\sigma}_0 \\ \hat{\sigma}_x \\ \hat{\sigma}_y \\ \hat{\sigma}_z \end{pmatrix}
%	,
%	\\ 
	\label{eq:3-2-23} %\ref{eq:3-2-23}
\tilde{\Gamma}^{\textrm{ra}}
	& = 
	\begin{pmatrix}
	&1 -i\Omega \tau - \frac{1}{2}\ell^2q^2 & \frac{i}{2} \ell q_y & -\frac{i}{2} \ell q_x  & 0\\
	& \frac{i}{2} \ell q_y 				
		& \frac{1}{2} ( 1 - i \Omega \tau  - \frac{1}{2}\ell^2  q^2) 
				 	& \frac{1}{4} \ell^2   q_{x}q_{y} & - \frac{i}{4} q_x \ell \frac{\hbar }{\epsilon_\textrm{F}\tau} \\
	 & -\frac{i}{2} \ell q_x
	 	& \frac{1}{4} \ell^2   q_{x}q_{y}
	 		&  \frac{1}{2} ( 1 - i \Omega \tau  - \frac{1}{2}\ell^2  q^2)  & - \frac{i}{4} q_y \ell \frac{\hbar }{\epsilon_\textrm{F}\tau} \\
	& 0 & \frac{i}{4} q_x \ell \frac{\hbar }{\epsilon_\textrm{F}\tau} & \frac{i}{4} q_y \ell \frac{\hbar }{\epsilon_\textrm{F}\tau} & o(\frac{\hbar }{\epsilon_\textrm{F}\tau})^2
	\end{pmatrix}.
\end{align}
In the above equation, we have used $n_{\rm{i}} u_{\rm{i}}^2 \pi\nu_e/ (\hbar /2\tau) =1/2 $.
Here, $\nu_e = \epsilon_{\textrm{F}}/(2\pi \hbar^2 \tilde{v}_\textrm{F}^2 )$ is the density of states at the Fermi energy on the surface of the doped TI.
From the above equation, the magnitudes of 
$\tilde{\Gamma}_{\zeta z}$ and $\tilde{\Gamma}_{z \zeta}$ are negligibly smaller than that of 
$\tilde{\Gamma}_{\nu \mu } (\nu, \mu =0, x, y)$ 
for $\hbar/(\epsilon_\textrm{F}\tau)\ll1$.
As a result, 
$\hat{\Gamma}_{\mu} = \sum_{\nu=0,x,y} [\tilde{\Gamma}]_{\mu\nu}\hat{\sigma}_\nu +o(\hbar/(\epsilon_\textrm{F}\tau))$ is obtained by 
\begin{align} \label{eq:3-2-24} %\ref{eq:3-2-24}
\hat{\Gamma}_{0}^{\textrm{ra}}
	= & \ \biggl(1- i\Omega \tau - \frac{1}{2}\ell^2 q^2\biggr) \hat{\sigma}_0 
	+ \frac{i}{2}\ell \hat{\sigma}_a q_b \epsilon_{ a b z}, 
	\\ 
	\label{eq:3-2-25} %\ref{eq:3-2-25}
\hat{\Gamma}_{\mu =x,y}^{\textrm{ra}}
	= &  \biggl[ \frac{1}{2}\biggl(1- i\Omega \tau - \frac{3}{4}\ell^2 q^2 \biggr)\delta_{\mu \nu} + \frac{1}{4}\ell^2 q_\mu q_\nu  \biggr] \hat{\sigma}_\nu
		 + \frac{i}{2}\ell q_a \epsilon_{\mu a z}\hat{\sigma}_0.
\end{align}
Then, $\tilde{\Lambda}^{\rm{ra}}_{\zeta\gamma} $ can also be estimated  by using $\tilde{\Gamma}^{\textrm{ra}}$ as\cite{ref:matrix}
\begin{align}
	\label{eq:3-2-26} %\ref{eq:3-2-26}
\tilde{\Lambda}^{\textrm{ra}}_{\gamma \zeta} & = [(1-\tilde{\Gamma}^{\textrm{ra}})^{-1}]_{\gamma \zeta}.
\end{align}
Therefore, from Eqs. (\ref{eq:3-2-13}) and (\ref{eq:3-2-21})-(\ref{eq:3-2-25}), the charge density $\rho_e$ is obtained by 
\begin{align} \notag
\rho_e 
	& = \frac{ e \nu_e J_{sd}\tau}{L^2} \sum_{\bm{q},\Omega}e^{i (\Omega t-\bm{q}\cdot\bm{x} )} 
	 \frac{ \ell \Omega }{q^2 \ell^2 + i\Omega \tau} (q_y S_{\bm{q},\Omega}^x - q_x S_{\bm{q},\Omega}^y)
	\\ 
	\label{eq:3-2-27} %\ref{eq:3-2-27}
	& =  - e \nu_e J_{sd} \tau \ell   [ \bm{\nabla} \times \partial_t \langle \bm{S}^\parallel\rangle_{\textrm{D}} ]_z,
\end{align}
From Eq. (\ref{eq:3-2-27}), 
we find that the charge density $\rho_e $ is induced by $\partial_t [\bm{\nabla} \times \langle \bm{S}^\parallel\rangle_{\textrm{D}} ]_z$.
Here $\langle \bm{S}^\parallel \rangle_{\rm{D}}$ is defined by the convolution of the in-plane of the localized spin $\bm{S}^\parallel$ and a 
diffusion propagator function $\mathcal{D}$ on the disordered surface of the TI as%\cite{rf:book-a}   
\begin{align}
	\label{eq:3-2-28} %\ref{eq:3-2-28}
\langle \bm{S}^\parallel \rangle_{\rm{D}} (\bm{x},t)
	& \equiv \frac{1}{\tau}\int dt' \int d\bm{x'}  \mathcal{D} (\bm{x}-\bm{x'},t-t') \bm{S}^\parallel (\bm{x'}, t'), 
	\\
	\label{eq:3-2-29} %\ref{eq:3-2-29}
\mathcal{D} (\bm{x},t)
	& = \frac{1}{L^2} \sum_{\bm{q},\Omega}e^{i(\Omega t-\bm{q}\cdot\bm{x})}  \frac{1}{ 2 D q^2  + i\Omega }, 
\end{align}
where, $ D \equiv \tilde{v}_\textrm{F} ^2 \tau/2$ is a diffusion constant and 
$\langle \bm{S}^\parallel \rangle_{\rm{D}}$ denotes the nonlocal spin, which diffusively propagates by the diffusion propagator $\mathcal{D}$.
The $\mathcal{D}$ results because of nonmagnetic impurity scattering on the disordered surface of the doped TI.
We also find that 
the charge density due to the out-of plane of the localized spin, $S^z$, is negligible smaller than that due to $\bm{S}^\parallel$.  

The diffusion propagator  $\mathcal{D}$ satisfies the differential equation%\cite{rf:book-a} 
\begin{align} 
	\label{eq:3-2-30} %\ref{eq:3-2-30}
( \partial_t  - 2 D \nabla^2 ) \mathcal{D}(\bm{x}-\bm{x'}, t-t') & = \delta(\bm{x}-\bm{x'}) \delta(t-t').
\end{align}
We find that from Eqs. (\ref{eq:3-2-27}) and (\ref{eq:3-2-28}), the diffusive motion of the charge density obeys the diffusion equation: 
\begin{align}
	\label{eq:3-2-31} %\ref{eq:3-2-31}
	( \partial_t  - 2 D \nabla^2 ) \rho_e  & =  - e \nu_e J_{sd} \ell ( \bm{\nabla} \times \partial_t \bm{S}^\parallel )_z.
\end{align}
The above equation means that the diffusion propagator of the charge density is caused by the spatial and time derivative of the localized spin, $( \bm{\nabla} \times \partial_t \bm{S}^\parallel )_z$, on the surface of the doped TI.
When the localized spin is spatially uniform, $\rho_e$ is not driven by the magnetization dynamics.

%%%%%%%%%%%%
\subsection{Spin current}
We will now consider the spin current due to the magnetization dynamics on the disordered surface of the doped TI.
The spin current is proportional to the charge density [see Eq. (\ref{eq:3-1-9})].
From the result of the charge density due to magnetization dynamics [see Eq. (\ref{eq:3-2-27})], the spin current is given by  
\begin{align} \label{eq:3-3-32} %\ref{eq:3-3-32}
j_{i}^\alpha  
	%	= &  \frac{v_{\rm{F}}}{2e} \epsilon_{z  i\alpha} \left[ - 2 e \nu_e J_{sd} \ell \tau \epsilon_{\nu \beta  z}  \partial_t \nabla_{\beta}  \langle S^\nu \rangle_{\rm{D}}\right]
%	\\
		= &   - \frac{1}{2} \epsilon_{z  \alpha i}  \nu_e J_{sd}  \ell^2    [ \bm{\nabla}\times  \partial_t \langle \bm{S}^\parallel \rangle_{\textrm{D}}]_z.
%	\\
%		= &  	 \nu_e J_{sd} \ell^2    \partial_t 
%		[ \nabla_{\alpha}  \langle S^i \rangle_{\rm{D}} - \nabla_{i}  \langle S^\alpha \rangle_{\rm{D}} ].
\end{align}
This is one of the main results of this paper.
From Eq. (\ref{eq:3-3-32}), the direction of spin of the spin current is perfectly locked and is perpendicular to the direction of the flow of the spin current.
The origin lies on the spin-momentum locking on the surface of the TI.
The spin current is proportional to the coefficients, which are the density of states at Fermi energy $\nu_e$, the $s$-$d$ exchange coupling $J_{sd}$, and the square of the mean-free path $\ell^2$. 
Here $j_{i}^\alpha$ is proportional to the spatial and time derivative of the nonlocal spin as $ [ \bm{\nabla} \times \partial_t \langle \bm{S}^\parallel \rangle_{\rm{D}} ]_z$.
We find that the local spin does not contribute to the spin current generation.
In the case when the spin structure of the MI is spatially uniform, the spin current vanishes.
Since the spin current is proportional to the charge density, 
we expect that the spin current can be arising from the accumulation of the diffusive charge density, which is given by Eq. (\ref{eq:3-2-31}).
%%%%%%%%%%%%%%%%%%%%%%%%%%%%%
Additionally, Eq. (\ref{eq:3-3-32}) indicates that  the spin current is an 
even-function of $\tilde{v}_\textrm{F} $, the  sign of which depends 
on the helicity of electron on the surface of the TI.
Therefore, the direction of the spin and flow of the spin current on top surface ($j^\alpha_{i,\textrm{top}}$) and that on bottom surface of the TI ($j^\alpha_{i,\textrm{bottom}}$) are equal as  $j^\alpha_{i,\textrm{top}} = j^\alpha_{i,\textrm{bottom}}$.
%%%%%%%%%%%%%%%%%%%%%%%%%%%%%
We find that the in-plane component of the localized spin $\bm{S}^\parallel \equiv \bm{S} - S^z \bm{z}$ contributes to the spin current, but the out-of plane component of the localized spin $S^z \bm{z}$ does not.
We expect that its origin lies on the spin-orbit coupling of $\mathcal{H}_{\textrm{TI}}$. 
From $(\hat{\bm{\sigma}}\times \bm{p})_z = \hat{\bm{\sigma}} \cdot (\bm{p} \times \bm{z})$ and $\bm{S}^\parallel = (\bm{z} \times \bm{S}^\parallel) \times \bm{z}$,
the Hamiltonian $\mathcal{H}_{\textrm{TI}} + \mathcal{H}_{sd}$ can be described by 
\begin{align} \label{eq:3-3-33} %\ref{eq:3-3-33}
\mathcal{H}_{\textrm{TI}} + \mathcal{H}_{sd} 
	& =  \int d\bm{x} \psi^\dagger 
		\biggl\{ 
		\tilde{v}_\textrm{F}  \hat{\bm{\sigma}} \cdot 
		\biggl[ \biggl( \bm{p}  - \frac{J_{sd}}{ \tilde{v}_\textrm{F}  } (\bm{z} \times \bm{S}^\parallel) \biggr) \times \bm{z}  \biggr] 
		- J_{sd} S^z \hat{\sigma}^z - \epsilon_\textrm{F} 
		\biggr\}\psi
\end{align}
From the above equation, we can regard that the conduction electrons momentum $\bm{p}$ is shifted by the in-plane localized spin $\bm{S}^\parallel$: 
$\bm{p} \to \bm{p} - \frac{J_{sd}}{\tilde{v}_\textrm{F} } (\bm{z} \times \bm{S}^\parallel)$.
The in-plane localized spin $\bm{z} \times \bm{S}^\parallel$ plays a role like an electromagnetic vector potential $\bm{\mathcal{A}} = \frac{J_{sd}}{e \tilde{v}_\textrm{F} } (\bm{z} \times \bm{S}^\parallel)$ \cite{rf:Nomura10,rf:Ueda12}.
%%%%%%%%%%%
Then, the observable quantity should be proportional to the gauge invariant form: 
an effective electric field $\bm{\mathcal{E}} \equiv -\partial_t \bm{\mathcal{A}} $ or an effective magnetic field $\bm{\mathcal{B}} = \bm{\nabla} \times  \bm{\mathcal{A}} $, as represented by 
\begin{align}
	\label{eq:3-3-34} %\ref{eq:3-3-34}
\bm{\mathcal{E}}
& = - \frac{J_{sd}}{e\tilde{v}_\textrm{F} } (\bm{z} \times \partial_t \bm{S}^\parallel), 
	\\
	\label{eq:3-3-35} %\ref{eq:3-3-35}
\bm{\mathcal{B}}
& = \frac{J_{sd}}{e\tilde{v}_\textrm{F}  } \bm{\nabla}\times (\bm{z} \times \bm{S}^\parallel).
\end{align}
The dynamics of the in-plane component of the localized spin 
can be regarded as the effective electromagnetic field, 
which acts as a driving force to trigger the motion of conduction electrons. 
While, the out-of plane one $S^z \bm{z}$ plays 
a role like magnetic fields for the conduction electrons and does not directly shift $\bm{p}$ in the momentum space. 
We expect from the difference of these properties of the localized spin, the contribution from $S^z \bm{z}$ could be smaller than that from $\bm{S}^\parallel$.

The spin current can be represented by using the effective electric field $\bm{\mathcal{E}}$ and $ \{\bm{\nabla}\times   [ \langle \bm{\mathcal{E}}\rangle_{\textrm{D}} \times \bm{z}] \}_z = -  \bm{\nabla} \cdot \langle \bm{\mathcal{E}}\rangle_{\textrm{D}} $ as 
\begin{align} \label{eq:3-3-36} %\ref{eq:3-3-36}
j_{i}^\alpha  
		& =  - \frac{1}{2} \epsilon_{z  \alpha i}  \nu_e e  \tilde{v}_\textrm{F}  \ell^2   \bm{\nabla} \cdot \langle \bm{\mathcal{E}}\rangle_{\textrm{D}} .
\end{align}
From the above equation, we find that the spin current is proportional to $\bm{\nabla} \cdot  \langle \bm{\mathcal{E}} \rangle_{\rm{D}}$  
stemming from  charge density.
In fact, the charge density can be represented by $\rho_e \propto \bm{\nabla} \cdot  \langle \bm{\mathcal{E}} \rangle_{\rm{D}}$, as shown in Eq. (\ref{eq:3-2-27}). 
Here, the charge density is also proportional to $\bm{\nabla} \cdot  \langle \bm{\mathcal{E}} \rangle_{\rm{D}}$ and is similar to the Gauss's law in Maxwell equations as $\rho_e =  \epsilon\bm{\nabla}\cdot \bm{E} $, 
where $\epsilon$ is a permittivity and $\bm{E}$ is an applied electric field.
Thus, we can interpret that  the charge density and the spin current on the surface of the TI  are generated by the divergence of the effective electric field.
Equations (\ref{eq:3-2-27}), (\ref{eq:3-3-32}), and (\ref{eq:3-3-36}) 
are the main results of this section.

%Charge current
\section{Charge current due to magnetization dynamics}
In this section, 
we show charge current due to magnetization dynamics on the disordered surface of the doped TI.
Because of the spin-momentum locking, the charge current is proportional to the density of the spin polarization on the surface of the doped TI.
We calculate the spin density, the charge current and  the resulting 
spin-relaxation torque.

%Spin density
\subsection{Spin density}
To discuss the charge current, we calculate the spin density due to the magnetization dynamics in the linear response to the localized spin.
The spin density $\bm{s} = \frac{1}{2}\langle \psi^\dagger \hat{\bm{\sigma}} \psi \rangle $ is given by 
\begin{align}
	\label{eq:4-1-37} %\ref{eq:4-1-37}
s^\mu 
	 =  \frac{i \hbar J_{sd} }{2L^2}\sum_{\bm{q},\Omega}e^{i (\Omega t-\bm{q}\cdot\bm{x} )} 
		{\rm{tr}}[ \hat{\Pi}_{\mu \nu} (\bm{q},\Omega) S^\nu_{\bm{q},\Omega} ], 
\end{align}
where, $\Pi_{\mu\nu} (\mu,\nu = x,y,z)$ is the spin-spin correlation function. 
$\Pi_{\mu\nu}$ can be  calculated within the same formalism as 
in the section 3.2,  
and is represented by 
$\hat{\Pi}_{\mu \nu} = \hat{\sigma}_\mu \hat{\Pi}_{0 \nu}$. 
From the result, we can obtain the spin density $\bm{s}$. 
Here, $\bm{s}$ can be decomposed into two terms: $\bm{s}= \bm{s}^\parallel + s^z \bm{z}$, where 
$\bm{s}^\parallel = \bm{s} - s^z \bm{z}$ and $s^z \bm{z}$ show the in-plane and out-of plane component of the spin on the disordered surface of the doped TI, respectively.
We find that  $s^z \bm{z}$ is proportional to $\partial_t S^z$, and its magnitude is negligibly smaller than that of the magnitude of $\bm{s}^\parallel$ within the approximation $|s^z |/ |\bm{s}^\parallel | \sim  o [ \hbar / (\epsilon_\textrm{F}\tau) \ll1 ]$.
Thus, the spin density can be estimated by $\bm{s} =  \bm{s}^\parallel + o [ \hbar / (\epsilon_\textrm{F}\tau) ]$ and $S^z$ does not contribute to the generation of $\bm{s}$.
The dominant contribution of $\bm{s}$ can also be decomposed into two terms:
\begin{align}
	\label{eq:4-1-38} %\ref{eq:4-1-38}
\bm{s} & =  \bm{s}^{\textrm{L}} + \bm{s}^{\textrm{D}},
\end{align}
where 
$\bm{s}^{\textrm{L}}$ is the local spin density due to $\bm{S}^\parallel$, and is given by 
\begin{align}
	\label{eq:4-1-39} %\ref{eq:4-1-39}
\bm{s}^{\textrm{L}} &=  - \frac{1}{2} \nu_e J_{sd} \tau \partial_t \bm{S}^\parallel .
\end{align}
The local spin density $\bm{s}^{\textrm{L}}$ is induced by 
the time-derivative of the in-plane component of the localized spin $\partial_t \bm{S}^\parallel$.
On the other hand, the second term of Eq. (\ref{eq:4-1-38}), $\bm{s}^{\textrm{D}}$, is the diffusive spin density and is given by
\begin{align}
	\label{eq:4-1-40} %\ref{eq:4-1-40}
\bm{s}^{\textrm{D}} 
	& = - \frac{1}{2}\nu_e J_{sd} \tau \ell^2 \left( \bm{z} \times \bm{\nabla} \right) [\bm{\nabla} \times \partial_t \langle \bm{S}^\parallel \rangle_{\textrm{D}} ]_z
	\\
	\label{eq:4-1-41} %\ref{eq:4-1-41}
	& = \frac{\ell}{2e} \left( \bm{z} \times \bm{\nabla} \right)  \rho_e.
\end{align}
From Eq. (\ref{eq:4-1-41}), $\bm{s}^{\textrm{D}}$ is generated by the driving field $(\bm{z} \times \bm{\nabla}) ( \bm{\nabla}\times \partial_t \langle  \bm{S} \rangle_{\textrm{D}} )_z   $, which is the spatial gradient and the time-derivative of the nonlocal localized spin $\langle \bm{S}^\parallel \rangle_{\textrm{D}}$.
The driving field is also described by $(\bm{z} \times \bm{\nabla}) ( \bm{\nabla}\times \partial_t \langle  \bm{S} \rangle_{\textrm{D}} )_z   
	 = 	\partial_t [\nabla^2  \langle  \bm{S} \rangle_{\textrm{D}} - \bm{\nabla} (\bm{\nabla}\cdot  \langle  \bm{S} \rangle_{\textrm{D}} )]$.
Here, $\bm{s}^{\textrm{D}}$ is described by the spatial gradient of the charge density, which is caused by the electron diffusion on the surface of the TI, as shown in Eq. (\ref{eq:3-2-27}).
In addition, $\bm{s}^{\textrm{D}}$ is also represented by the spin current: 
The charge density is proportional to the spin current, $\rho_e  =  \frac{e} {\tilde{v}_\textrm{F}} (j_y^x  -j_x^y )$, and $\bm{s}^{\textrm{D}}$ becomes 
\begin{align}
\bm{s}^\textrm{D}	
	& = \frac{1}{2} \tau \epsilon_{z \alpha i} (\bm{z} \times \bm{\nabla})   j_i^\alpha  .
\end{align}
From Eqs. (\ref{eq:4-1-39})-(\ref{eq:4-1-40}), we find that $\bm{s}$ is an 
even-function of $\tilde{v}_\textrm{F} $ and is independent of the helicity on the surface of the TI.
Therefore, the direction of $\bm{s}$ does not depend on 
whether we are focusing on the top or bottom surface of the TI.

%Discuss and comment from A. Yamakage
We find that from Eqs. (\ref{eq:4-1-39})-(\ref{eq:4-1-40}), the spin is polarized not by a static magnetization but by magnetization dynamics.
Therefore, we expect that static magnetization does not induce spin polarization on the surface of the TI. 
This seems to be anomalous property on the surface.
The response between the spin polarization and the static magnetization on the surface of the TI is different from that in conventional metals: In the metals, a spin is polarized even by static magnetization.
We will consider shortly why static magnetization does not generate spin polarization on the surface of the TI.
The magnetization on the surface of the TI plays the role to shift the momentum of conduction electrons from $\bm{p}$ into $\bm{p} - \frac{J_{sd}}{\tilde{v}_F}(\bm{z}\times \bm{S}^\parallel)$ in momentum space.
As a result, the center of Fermi sphere is also shifted from $\bm{p}=0$ into $\bm{p}=\frac{J_{sd}}{\tilde{v}_F}(\bm{z}\times \bm{S}^\parallel)$. 
Then, the direction of the spin at each momentum are perfectly perpendicular to that of the momentum.
Besides, the spin configuration in momentum space does not change before and after the shift, because the direction of the spin at each momentum are independent of the shift.
Therefore, no spin polarization is driven by momentum shift due to a static magnetization.

\subsection{Charge current}
We note that on the surface of the TI, the charge current $\bm{j}$ 
is proportional to the spin density 
on the surface of the TI. The charge current is represented by using the renormarized velocity operator as 
$\bm{j} =  2 e \tilde{v}_\textrm{F}  (\bm{z} \times \bm{s})$\cite{rf:discussion1}. 
From Eqs. (\ref{eq:4-1-39})-(\ref{eq:4-1-41}), the charge current is also decomposed into two terms: 
$\bm{j} = \bm{j}^{\textrm{L}} + \bm{j}^{\textrm{D}}$ as 
\begin{align}
	\label{eq:4-2-43} %\ref{eq:4-2-43}
\bm{j}^{\textrm{L}}
	& = 	-  e \nu_e J_{sd} \ell ( \bm{z} \times \partial_t \bm{S}^\parallel ),
	\\
	\label{eq:4-2-44} %\ref{eq:4-2-44}
\bm{j}^{\textrm{D}}
	& =   e \nu_e J_{sd}  \ell^3 \bm{\nabla}  [\bm{\nabla} \times \partial_t \langle \bm{S}^\parallel \rangle_{\textrm{D}} ]_z.
\end{align}
The $\bm{j}^{\textrm{L}}$ is the local charge current and is induced by the 
time-derivative of the localized spin $\bm{S}^\parallel$. The direction of $\bm{j}^{\textrm{L}}$ is parallel to $\bm{z} \times \bm{S}^\parallel$. 
On the other hand, 
$\bm{j}^{\textrm{D}}$ is the diffusive charge current caused by impurity scattering on the disordered surface of the TI.
In fact, $\bm{j}^{\textrm{D}}$ can be represented by the spatial gradient of the charge density as $\bm{j}^{\textrm{D}}  = - 2D \bm{\nabla} \rho_e$. This means that  diffusive current is generated by  the spatial gradient of the charge density on the surface of the TI.
We note that the charge current is an odd-function of $\tilde{v}_\textrm{F} $, so that the direction of the charge current on the top surface $\bm{j}_{\textrm{top}}$ is opposite to that on the bottom surface $\bm{j}_{\textrm{bottom}}$, if $\bm{S}$ is same on the top and bottom surface.
It is noted that, from Eqs. (\ref{eq:3-2-27}) and (\ref{eq:4-2-43})-(\ref{eq:4-2-44}), the charge density $\rho_e$ and charge current $\bm{j}$ satisfy the conservation law $\dot{\rho}_e+ \bm{\nabla}\cdot \bm{j}_\gamma =0$. The detail is shown in Appendix \ref{sec:Charge conservation}.

Next, we comment on the relationship between the spin current and the charge current on the disordered surface of the doped TI. 
Substituting $\epsilon_{z \alpha i} j_{i}^\alpha = \tilde{v}_\textrm{F}  \rho_e /e$ into Eq. (\ref{eq:4-2-44}), we find that the diffusive charge current can be described by the spin current as 
\begin{align}
		\label{eq:4-2-45} %\ref{eq:4-2-45}
\bm{j}^{\textrm{D}}
%	& =  - \frac{eD}{v_\textrm{F}} \bm{\nabla} \epsilon_{z \alpha i} j_{i}^\alpha
%	\\
	& =  - e\ell  \epsilon_{z \alpha i} \bm{\nabla} j_{i}^\alpha.
\end{align}
This is also the main result of this paper.
We expect that the above equation displays the conversion between the spin current into the diffusive charge current on the disordered surface of the doped TI by using the spatial gradient of the spin current. 
The spin current can be converted into the diffusive charge current when the spin current depends on the space on the disordered surface.
The relation in Eq. (\ref{eq:4-2-45}) is plausible on the disordered surface of the doped TI, 
because the charge density $\rho_e$ is proportional to the spin current, and a diffusive particle current generally proportional to a spatial gradient of particles. % $\bm{j}^\textrm{D} \propto  -D \bm{\nabla} \rho_e$.
We note that there is no relation between the spin current and the local charge current $\bm{j}^\textrm{L}$.

\subsection{Effective conductivity}
The charge current due to magnetization dynamics $\bm{j}$ can be also described by the effective electric field $\bm{\mathcal{E}}$:
\begin{align}
		\label{eq:4-3-46} %\ref{eq:4-3-46}
\bm{j}
	& = e^2 \tilde{v}_\textrm{F}^2   \nu_e  \tau   \bm{\mathcal{E}}
		+  e^2 \tilde{v}_\textrm{F} \nu_e   \ell^3 \bm{\nabla} [ \bm{\nabla}\cdot \langle  \bm{\mathcal{E}} \rangle_D].
\end{align}
The first term and the second term are corresponding to the local and diffusive charge current, respectively.
From the above results, we will consider an effective conductivity: an efficiency of the charge flow due to the applied effective electric field $\bm{\mathcal{E}}$. 
This is similar to the conventional electric conductivity: 
In general, the longitudinal electrical  conductivity is defined from dividing the charge current by an applied electric field\cite{rf:Kittel}.
We expect that the current corresponds to the local current.
Then, an effective longitudinal electrical conductivity can be defined by $\bm{j}^{\textrm{L}} = \sigma \bm{\mathcal{E}}$, and is given by the first term in Eq. (\ref{eq:4-3-46}) as 
\begin{align}
		\label{eq:4-3-47} %\ref{eq:4-3-47}
\sigma & = e^2 \tilde{v}^2_{\rm{F}} \nu_e \tau.  
\end{align}
The conductivity only depends on the parameters on the surface of the TI, and is independent of parameters attached to the MI.

\subsection{Spin relaxation torque}
We will consider the spin relaxation due to the magnetization dynamics on the surface of the TI. 
Using Eqs. (\ref{eq:3-3-32}) and (\ref{eq:4-1-38})-(\ref{eq:4-1-40}), we can describe $\partial_t s^\alpha$ and $\nabla_i j_i^{\alpha}$as  
\begin{align}
	\label{eq:4-4-48} %\ref{eq:4-4-48}
\partial_t s^\alpha & = - \frac{1}{2} \nu_e J_{sd}\tau \partial_t^2 S^{\parallel, \alpha}  
				 + \frac{1}{2} \nu_e J_{sd}  \ell^2 \tau  \epsilon_{\alpha iz } \nabla_i [\bm{\nabla} \times \partial_t^2 \langle \bm{S}^\parallel \rangle_{\textrm{D}} ]_z, 	\\
	\label{eq:4-4-49} %\ref{eq:4-4-49}
\nabla_i j_i^{\alpha}
	& = - \frac{1}{2} \nu_e J_{sd}  \ell^2  \epsilon_{\alpha  i z   }  \nabla_i  [ \bm{\nabla}\times  \partial_t \langle \bm{S}^\parallel \rangle_{\textrm{D}}]_z.
\end{align}
Therefore, 
the spin relaxation torque $\mathcal{T}^\alpha = \partial_t s^\alpha + \nabla_i j_i^{\alpha}$ in the linear response to $\bm{S}^\parallel$ is obtained by
\begin{align} \label{eq:4-4-50} %\ref{eq:4-4-50}
\bm{\mathcal{T}}	
	= & - \frac{1}{2} \nu_e J_{sd}\tau \partial_t^2 \bm{S}^{\parallel}  
		 +  \frac{1}{2} \nu_e J_{sd}  \ell^2 (1 -  \tau \partial_t)  ( \bm{z} \times \bm{\nabla}) 
		   [ \bm{\nabla} \times \partial_t \langle  \bm{S}^\parallel \rangle_{\textrm{D}}]_z 
		   + \mathcal{O}(J_{sd}^2).		 
\end{align}
The first term in Eq. (\ref{eq:4-4-48}) shows a local spin relaxation torque and is induced by $\partial_t^2 \bm{S}^{\parallel}$. 
The second term is a nonlocal spin relaxation torque and is driven by $(\bm{z} \times \bm{\nabla} )( \bm{\nabla} \times \partial_t \langle  \bm{S}^\parallel \rangle_{\textrm{D}} )_z$.
The nonlocal torque vanishes when the magnetization is spatially uniform.
The third term in Eq. (\ref{eq:4-4-48}) represents the higher order of $J_\textrm{sd}$ and $\bm{S}$.
From Eq. (\ref{eq:3-1-12}), the spin relaxation torque $\bm{\mathcal{T}}_{sd}$ is proportional to $\bm{S}$ and $\bm{s}$, which is proportional to $\bm{S}^\parallel$.
Therefore, the third term in Eq. (\ref{eq:4-4-48}) corresponds to  $\bm{\mathcal{T}}_{sd}$ within the linear response to $\bm{S}$. 
We expect that 
the third term $\bm{\mathcal{T}}_{sd} \simeq o(\frac{J_{sd}\tau}{\hbar })^2 $ can be negligibly small and be ignored in comparison with the first and the second terms of $\mathcal{T}$ in the regime $\frac{J_{sd}\tau}{\hbar }\ll1$.

The spin relaxation torque $\bm{\mathcal{T}}$ is also represented by the effective electric field as 
\begin{align} \label{eq:4-4-51} %\ref{eq:4-4-51}
\bm{\mathcal{T}}	
	= &  \frac{1}{2} e \nu_e \tilde{v}_\textrm{F} \tau  (\partial_t \bm{\mathcal{E}}\times \bm{z})
		 + \frac{1}{2} e  \nu_e \tilde{v}_\textrm{F}   \ell^2 (1 -  \tau \partial_t) ( \bm{z} \times \bm{\nabla}) [ \bm{\nabla}\cdot \langle  \bm{\mathcal{E}} \rangle_{\textrm{D}}].
\end{align}
The local spin relaxation is written as the time-derivative of the effective electric field and 
the diffusive one is induced by the spatial gradient of the nonlocal effective electric field. 
The electric field dependence of the spin relaxation torque on the surface of the TI is different from that in NM with spin-orbit interactions: The spin relaxation torque in the NM, $\mathcal{T}_{\textrm{NM}}$, is proportional to the spatial gradient of the applied electric field\cite{rf:Nakabayashi}.
We expect that the difference can be caused by the $\bm{k}$-dependence of the energy dispersion:
The energy dispersion on the surface of the TI is a linear function 
of $\bm{k}$, while that in the NM proportional to the square of $\bm{k}$.
Equations (\ref{eq:4-1-40}), (\ref{eq:4-2-45}), and (\ref{eq:4-3-47}) are the main results of this section.

\section{Discussion}
\subsection{Spin torque}
We will phenomenologically study the change of the magnetization dynamics in before and after the spin-charge generation due to the ferromagnetic resonance (FMR).
Now, we consider a disordered surface of the MI/TI junction, as shown 
in Fig.1.
Further, in the junction, a static magnetic field and ac magnetic field are additionally applied.
Here ac magnetic field is given by a microwave irradiation, and is needed for 
FMR in the MI.  
When we apply this magnetic field, the magnetization dynamics is triggered in the MI, and the magnetization dynamics induces the spin polarization on the surface of the TI [see Eqs. (\ref{eq:4-1-38})-(\ref{eq:4-1-40})].
Then, the induced spin polarization $\bm{s}$ plays the role of an exchange field acting on the magnetization in the MI.
As a result, the magnetization dynamics is affected from the generated spin $\bm{s}$.
Here, the mutual interaction between the magnetization dynamics and the induced spin are called as the feedback effect\cite{rf:sakai14}.
%The magnetization dynamics induce spin polarization and its spin modifies the magnetization dynamics, and the modified magnetization drives spin polarization, .....

The magnetization dynamics on the disordered surface of the doped TI is described from the Landau--Lifshitz--Gilbert (LLG) equation as\cite{rf:Gilbert04}  
\begin{align} \label{eq:5-1-52} %\ref{eq:5-1-52}
\partial_t \bm{M} & =  - \gamma \mu  (\bm{M} \times \bm{H} ) + \frac{\alpha}{M} \bm{M} \times \partial_t  \bm{M} + \bm{\mathcal{T}}_{\rm{e}}, 
\end{align} 
where, $\bm{M}=- \frac{M}{S} \bm{S}$ is the magnetization of the MI, 
$\gamma$ is the gyromagnetic ratio, 
$\mu$ is a permeability,
$\alpha$ is a Gilbert damping constant, 
$\bm{H} = \bm{H}_0 + \bm{H}_{ac}$ is an applied magnetic field on the disordered surface of the doped TI.
$\bm{H}_0$ and $\bm{H}_{ac}$ denote a static and ac magnetic field, respectively.
The spin torque on the disordered surface of the doped TI is 
given by $\bm{\mathcal{T}}_{\rm{e}} = \frac{2 J_{sd}a^2 }{\hbar}  \bm{M} \times  \bm{s}$. 
The torque can be decomposed into two terms: $\bm{\mathcal{T}}_{\rm{e}} = \bm{\mathcal{T}}^\textrm{L}_{\rm{e}} + \bm{\mathcal{T}}^\textrm{D}_{\rm{e}}$, 
where $\bm{\mathcal{T}}^\textrm{L}_{\rm{e}} = \frac{2 J_{sd}a^2 }{\hbar} \bm{M} \times  \bm{s}^\textrm{L} $ and $\bm{\mathcal{T}}^\textrm{D}_{\rm{e}} = \frac{2 J_{sd}a^2}{\hbar} \bm{M} \times  \bm{s}^\textrm{D}  $ are the local and diffusive spin torque, respectively. Here, $a$ is a lattice constant on the surface of the TI. 
%
%\begin{align} 
%        \label{eq:5-1-53} %\ref{eq:5-1-53}
%\bm{\mathcal{T}}_{\rm{e}}
%         = & \frac{ \kappa}{M^\parallel} \biggl[  \bm{M} \times \partial_t \bm{M}^\parallel +
%   \ell^2  \bm{M}  \times   \left( \bm{z} \times \bm{\nabla} \right)   ( \bm{\nabla} \times \partial_t  \langle \bm{M^\parallel}\rangle_{\textrm{D}} )_z  \biggr],
%\end{align}
%
These spin torques are obtained from Eqs. (\ref{eq:4-1-38})-(\ref{eq:4-1-40}) and $\bm{S^\parallel}  = - \frac{S^\parallel}{M^\parallel} \bm{M}^\parallel$ as
\begin{align} 
        \label{eq:5-1-53} %\ref{eq:5-1-53}
\bm{\mathcal{T}}_{\rm{e}}^\textrm{L}
         = & \frac{ \kappa}{M^\parallel}   \bm{M} \times \partial_t \bm{M}^\parallel,  
        \\
	\label{eq:5-1-54} %\ref{eq:5-1-54}
\bm{\mathcal{T}}_{\rm{e}}^\textrm{D}
         = &  \frac{\kappa}{M^\parallel} \ell^2  \bm{M}  \times   \left( \bm{z} \times \bm{\nabla} \right)   ( \bm{\nabla} \times \partial_t  \langle \bm{M^\parallel}\rangle_{\textrm{D}} )_z  ,
\end{align}
where, $M^\parallel$ is the magnitude of the in-plane magnetization $\bm{M}^\parallel \equiv \bm{M} - M_z \bm{z}$, and $\kappa = \nu_e a^2 J_{sd}^2 \tau S^\parallel / \hbar $ is dimensionless coefficient proportional to $J_{sd}^2$ and $\tau$. 
%$\bm{\mathcal{T}}_{\rm{e}}^\textrm{L}$ is caused by $\bm{M} \times \bm{s}^{\parallel, \textrm{L}}$.
%
We find that $\bm{\mathcal{T}}_{\rm{e}}^\textrm{L} \propto  \bm{M} \times \partial_t \bm{M}^\parallel $ is slightly different from the damping torque $\frac{\alpha}{M} \bm{M}\times \partial_t \bm{M}$; which 
is a damping of the magnetization.
We could expect that the contribution from the local spin torque $ \bm{\mathcal{T}}_{\rm{e}}^\textrm{L}$ can be observed in the experiments on the surface of FM/TI junction\cite{rf:Deorani,rf:Shiomi14}. 
Here, $ \bm{\mathcal{T}}_{\rm{e}}^\textrm{L}$ plays the role of an anisotropic damping torque 
unless the static magnetic field and microwave are parallel 
to the $z$ direction.
The anisotropic damping affects the magnetic permeability:
For example, when the static magnetic field and microwave are parallel 
to the $y$-direction, 
then, the longitudinal magnetic permeability $\chi_{xx}$ and $\chi_{zz}$ are not equal each other.
%
%That is caused by the anisotropic damping.
Here, $\bm{\mathcal{T}}_{\rm{e}}^\textrm{D}$ seems to be a new type of spin torque $\bm{\mathcal{T}}_{\rm{e}}^\textrm{D}$ on the surface of the TI.
$\bm{\mathcal{T}}_{\rm{e}}^\textrm{D}$ in Eq. (\ref{eq:5-1-54}) is induced by the spatial gradient of the magnetization, $\bm{M}  \times   \left( \bm{z} \times \bm{\nabla} \right)   ( \bm{\nabla} \times \partial_t  \langle \bm{M^\parallel}\rangle_{\textrm{D}} )_z$.
When the magnetization is spatially uniform, $\bm{\mathcal{T}}_{\rm{e}}^\textrm{D}$ is zero and  $\bm{\mathcal{T}}_{\rm{e}}^\textrm{L}$ is nonzero.

Since, $j_{i}^\alpha$ and $\bm{\mathcal{T}}^{\textrm{D}}_{\textrm{e}}$ are proportional to the charge density on the disordered surface of the doped TI, 
$\bm{\mathcal{T}}^{\textrm{D}}_{\textrm{e}}$ can be described by the spin current $j_{i}^\alpha$:
\begin{align}
\label{eq:5-1-55} %\ref{eq:5-1-55}
\bm{\mathcal{T}}^{\textrm{D}}_{\textrm{e}}
	& =  \frac{J_{sd}\tau a^2}{\hbar}  \bm{M} \times [ (\bm{z} \times \bm{\nabla})  \epsilon_{z \alpha i} j_i^\alpha].
\end{align}
The spatial gradient of the spin current induces the diffusive spin torque on the disordered surface of the doped TI. 
We expect that the contribution of $j_{i}^\alpha$ is detected from the change of the half-width value, as well as the change of a shift of the magnetic resonance frequency through $\bm{\mathcal{T}}^{\textrm{D}}_{\textrm{e}}$ [discussed in Sec.\ref{sec:5-c}]. 
Equations (\ref{eq:5-1-53})-(\ref{eq:5-1-55}) are the main results of this section.

\subsection{Magnetic permeability without diffusion}
%%%%%%%%%%%%%% FIG2
%\begin{figure}[htbp]\centering \label{fig:1}%\ref{fig:2}
%\includegraphics[scale=.5]{fig2-TI.eps} 
%\caption{(Color online) 
%Schematic illustration of the spin-pumping due to the ferromagnetic resonance, where the static magnetic field and the ac magnetic field due to microwave irradiate are along the $y$-direction. }
%\end{figure}
%%%%%%%%%%%%
Using Eqs. (\ref{eq:5-1-52})-(\ref{eq:5-1-54}), 
we discuss the magnetic permeability in FMR when the magnetization is spatially uniform  on the surface of the MI/TI junction.
We consider that in the junction, the applied static magnetic field $\bm{H}_{0}$ and the microwave of the ac magnetic field $\bm{H}_{ac}$ are applied along the $y$ direction: $\bm{H}_{0} = (0, H, 0)$ and $\bm{H}_{ac} = (h_x, 0,h_z)$.
For an uniform magnetization case, 
the spin becomes $\bm{s}^\textrm{L}\neq0$ and $\bm{s}^\textrm{D}=0$, and the spin torque are $\bm{\mathcal{T}}_{\rm{e}}^\textrm{L}\neq 0$, but $\bm{\mathcal{T}}_{\rm{e}}^\textrm{D}=0$.  

Then, the LLG equation on the surface of the MI/TI junction can be described by
\begin{align} \label{eq:5-2-56} %\ref{eq:5-2-56}
\partial_t \bm{M}  =&   \gamma  \mu  (\bm{H} \times \bm{M})  + \frac{\alpha}{M}  (\bm{M} \times \partial_t  \bm{M})
	 	+ \frac{\kappa}{M^\parallel}  (\bm{M} \times \partial_t \bm{M}^\parallel).
\end{align} 
To estimate the magnetic permeability on the surface of the TI, 
we assume that the $|\bm{H}_0|$ is larger than the $|\bm{H}_{ac}|$, $i.e.$, 
$\mid h_x \mid \ll H $ and $\mid h_z \mid \ll H$.
Then, from the applied magnetic field, 
we expect that the local magnetization on the surface $\bm{M}= (m_x, M_y, m_z)$ can be satisfied $m_x \ll M_y$ and $m_z \ll M_y$.
Moreover we assume that the time-dependence of the precession of $\bm{M}$ is given by
$ m_x \propto m_z \propto e^{i\Omega t} $ and $\partial_t M_y \sim 0$.
In order to solve the LLG equation, we take a linear approximation of $m_i$: $m_i m_j \sim 0 $, $M_y \sim  M (=|\bm{M}|)$, and $M^\parallel \sim M$.
Then, the LLG equation becomes 
\begin{align} 
 \begin{matrix}
\partial_t m_x   =   \gamma \mu  (H m_z- h_z M ) 
	 +  \alpha \partial_t m_z,
	\\
	\label{eq:5-2-57} %\ref{eq:5-2-57}
\partial_t m_z   =   \gamma \mu  (h_x M - H m_x ) 
	  - (\alpha+\kappa)  \partial_t m_x,
 \end{matrix}
\end{align} 
and the magnetic permeability is given by 
\begin{align}\label{eq:5-2-58} %\ref{eq:5-2-58}
\biggl(\begin{matrix}
m_x \\ m_z
\end{matrix}\biggr)
 = & 
\biggl(\begin{matrix}
\chi_{xx} &  \chi_{xz} 
\\ 
\chi_{zx} & \chi_{zz}  
\end{matrix}\biggr)	
\biggl(\begin{matrix}
h_x \\ h_z
\end{matrix}\biggr).
\end{align} 
The frequency dependence of the longitudinal magnetic permeability $\chi_{xx}$ and $\chi_{zz}$ are described by 
\begin{align}
\label{eq:5-2-59} %\ref{eq:5-2-59}
\chi_{xx} 
	= & \frac{ \omega_M (\omega_H + i \alpha \Omega) }
	{( \omega_H + i\alpha \Omega  ) [ \omega_H + i (\alpha + \kappa) \Omega ] -\Omega^2}, 
	\\ 
	\label{eq:5-2-60} %\ref{eq:5-2-60}
\chi_{zz} 
	 = & \frac{  \omega_H + i (\alpha + \kappa) \Omega  }{ \omega_H + i \alpha  \Omega } \chi_{xx},
\end{align} 
where. $\omega_H = \gamma \mu H $ and $\omega_M = \gamma \mu M$ are the angular frequency of the applied static magnetic field 
and that of the magnetization, respectively.
The permeability $\chi_{xx}$ and $\chi_{zz}$ are different  each other 
originating from the anisotropic spin-transfer torque $\bm{\mathcal{T}}_e^{\textrm{L}}$.
Transverse magnetic permeability has the relation $\chi_{xz} = -\chi_{zx}$ and is given by 
\begin{align}
	\label{eq:5-2-61} %\ref{eq:5-2-61}
\chi_{xz}   
	= &   \frac{ -i \Omega  }{  \omega_H + i \alpha  \Omega  }\chi_{xx}.
\end{align} 
The real part of the permeability Re$[\chi_{xx}]$ shows a Lorentzian profile due to the magnetic resonance  around the resonant frequency $\Omega_r$, which is proportional to $H$.
The Im$[\chi_{xx}]$ indicates the energy absorption of the applied microwave around $\Omega_r$.
Half-width value of Im$[\chi_{xx}]$ expresses the damping of the precessional motion of the magnetization.
From Eq. (\ref{eq:5-2-59}), the half-width value $\Delta \Omega$ is caused by the Gilbert damping ($\alpha \bm{M}\times \partial_t \bm{M}$) and the local spin torque $\bm{\mathcal{T}}_e^\textrm{L}$ given by  
\begin{align}
	\label{eq:5-2-62} %\ref{eq:5-2-62}
\Delta \Omega & \simeq (2\alpha  + \kappa) \Omega_r.
\end{align}
We expect that 
in the MI without TI, the half-width value of the magnetic permeability estimates $\Delta \Omega  = 2\alpha \Omega_r $.
Eq. (\ref{eq:5-2-62}) indicates that on the surface of the doped TI, $\Delta \Omega$ is enhanced from $2\alpha \Omega_r$ into $(2\alpha + \kappa)\Omega_r$. 
The origin lies on $\bm{\mathcal{T}}_e^\textrm{L}$, which is triggered by the induced spin, 
where the spin is induced by the magnetization dynamics on the surface of the doped TI.
This enhancement of $\Delta\Omega$ has been verified in the recent 
experiment\cite{rf:Shiomi14}.

We compare $\Delta\Omega$ in MI/TI junction with $\Delta\Omega$ in FM/NM junction. 
In the FM/NM junction,  
it has been demonstrated that the enhancement of $\Delta \Omega$ is triggered by the spin current in the NM, which is generated by the magnetization dynamics of the FM\cite{rf:Mizukami02}.
On the surface of the MI/TI junction with uniform spin structure, 
on the other hand, 
spin current is not generated by magnetization dynamics, and the spin current does not contribute to $\Delta \Omega$.
The enhancement of $\Delta \Omega$ is caused by the spin polarization due to the magnetization dynamics on the surface.

%%%%%%%%%%%%%%
%% 	Discussion part 2  %%
%%%%%%%%%%%%%%
\subsection{Magnetic permeability with diffusion}\label{sec:5-c}
We will consider the surface of the TI/MI junction, where the localized spin in the MI is spatially inhomogeneous.
The spin structure of the localized spin we consider is a spin-wave or longitudinal conical spin structure, which are realized in ferrimagnetic insulator yttrium iron garnets (YIG) or multiferroics\cite{rf:Tokura11}, respectively.
%$e.g.,$ Y$_3$Fe$_5$O$_{12}$\cite{rf:Uchida10,rf:Kajiwara}.%\cite{rf:Uchida10,rf:Kajiwara}.
We expect that even on the surface of the TI, the localized spin depends on the position through proximity effects from the MI.
Then, if we apply ac magnetic field of microwave in the junction along the $y$ direction, 
we assume that the localized spin becomes precessional motion by the applied magnetic field.
The localized spin on the surface of the TI can be described by
\begin{align} 
	\label{eq:5-3-63} %\ref{eq:5-3-63}
\bm{S}(\bm{x},t) & = [S \cos{(\bm{q}\cdot\bm{x} -\Omega t)}, S_y, S \sin{(\bm{q}\cdot\bm{x} -\Omega t)}], 
\end{align} 
where, $S$ and $S_y$ are a constant coefficient independent of space. 
We assume $\mid S \mid \ll \mid S_y \mid$ and $\mid\bm{S}\mid \sim S_y$.
Here $\bm{q} = (q_x, q_y)$ is the momentum of the localized spin and is assumed to be  monochromatic.
The direction and the magnitude of $\bm{q}$ depends on materials of the MI.
In the spin structure, nonlocal diffusive spin $\langle \bm{S} \rangle_{\textrm{D}}$ is given by Eqs. (\ref{eq:3-2-28}) and (\ref{eq:5-3-63}) as
\begin{align} 	
	\label{eq:5-3-64} %\ref{eq:5-3-64}
	\begin{matrix}
\langle S_x \rangle_\textrm{D} 
		 = 
		A_{q,\Omega} S \cos{[ \bm{q} \cdot \bm{x}  -\Omega t ]} 
		-
		B_{q,\Omega} S \sin{[ \bm{q} \cdot \bm{x}  -\Omega t ]},
		\\
\langle S_z \rangle_\textrm{D} 
		 = 
		B_{q,\Omega} S \cos{[ \bm{q} \cdot \bm{x}  -\Omega t ]} 
		+ 
		A_{q,\Omega} S \sin{[ \bm{q} \cdot \bm{x}  -\Omega t ]}.
	\end{matrix}
\end{align}
The component of $\langle \bm{S} \rangle_{\textrm{D}}$ is different from that of $\bm{S}$: 
$\langle S_x \rangle_{\textrm{D}}$ has $\cos{[ \bm{q} \cdot \bm{x}  -\Omega t ]}$ components, as well as $\sin{[ \bm{q} \cdot \bm{x}  -\Omega t ]}$ components. 
Here, the coefficients $A$ and $B$ are obtained (see Appendix \ref{sec:A}) as 
\begin{align}
	\label{eq:5-3-65} %\ref{eq:5-3-65}
A_{q,\Omega}
	& = \frac{q^2 \ell^2}{ (\Omega \tau)^2 + (q^2 \ell^2)^2},
	\\
	\label{eq:5-3-66} %\ref{eq:5-3-66}
B_{q,\Omega}
	& = \frac{\Omega \tau}{ (\Omega \tau)^2 + (q^2 \ell^2)^2}.
\end{align}
The coefficients $A_{q,\Omega}$ and $B_{q,\Omega}$ depend on $q\ell$ and $\Omega \tau$, where 
parameters $\ell $ and $\tau$ are determined by the TI, and 
$q$ and $\Omega$ are chosen as characteristic values of the MI.
Then, the diffusive spin $\bm{s}^\textrm{D}$ is given from Eqs. (\ref{eq:4-1-40}) as %and by using 
%$(\bm{z} \times \bm{\nabla}) ( \bm{\nabla} \times \partial_t \langle  \bm{S}^\parallel \rangle_{\textrm{D}} )_z
%		= -(\delta_{\alpha \beta}\delta_{i\ell}- \delta_{\alpha \ell}\delta_{\beta i})
%		 \partial_i \partial_\beta \partial_t \langle S^{\parallel,\ell} \rangle_{\textrm{D}}$ as 
%
\begin{align}
	\label{eq:5-3-67} %\ref{eq:5-3-67}
\bm{s}^\textrm{D}	
	& =  \nu_e J_{sd}\tau \ell^2  [q^2 \partial_t \langle  \bm{S^\parallel} \rangle_{\textrm{D}} - \bm{q} (\bm{q} \cdot \partial_t \langle  \bm{S^\parallel} \rangle_{\textrm{D}} ) ].
\end{align}
Then, nonlocal diffusive spin torque $\bm{\mathcal{T}}^{\textrm{D}}_{\textrm{e}}$ is obtained by using $ \langle \bm{M}^\parallel \rangle_{\textrm{D}} = - (M^\parallel/S^\parallel)  \langle \bm{S}^\parallel \rangle_{\textrm{D}} $ as 
\begin{align}
\label{eq:5-3-68} %\ref{eq:5-3-68}
\bm{\mathcal{T}}^{\textrm{D}}_{\textrm{e}}
	& = - \frac{\kappa \ell^2}{M^\parallel}  
		\bigl[q^2  \bm{M} \times \partial_t  \langle \bm{M}^\parallel  \rangle_{\textrm{D}}
		- (\bm{M}\times \bm{q}) (\bm{q} \cdot \partial_t  \langle \bm{M}^\parallel  \rangle_{\textrm{D}}) 
		 \bigr].
\end{align}
The first term of Eq. (\ref{eq:5-3-68}) plays the role of an anisotropic damping torque, which is similar to $\mathcal{T}^{\textrm{L}}_{\textrm{e}}$. 
The direction of this torque is independent of the direction of $\bm{q}$. 
On the other hand, the second term of Eq. (\ref{eq:5-3-68}) is proportional to $(\bm{M}\times \bm{q}) (\bm{q} \cdot \partial_t  \langle \bm{M}^\parallel  \rangle_{\textrm{D}}) $, 
and its direction depends on $\bm{q}$. 

Using Eqs. (\ref{eq:5-3-63})-(\ref{eq:5-3-68}), we consider 
the magnetic permeability affected by $\bm{\mathcal{T}}^{\textrm{D}}_{\textrm{e}}$.%, when $\bm{q}= q \bm{x}$ is satisfied in the MI. 
Then, the LLG equation is given 
within the linear order of the magnetization as 
\begin{align*}
% X-components 
	\partial_t m_x   & =   \omega_H m_z - h_z \omega_M +   \alpha \partial_t  m_z 
	 \\  
% Z-components 
	\partial_t m_z  & = h_x \omega_M - \omega_H m_x  - (\alpha + \kappa)\partial_t  m_x 
				    + \kappa q_y^2 \ell^2 \partial_t \langle M^{\parallel, x} \rangle_{\textrm{D}}.
\end{align*} 
The last term of the above equation is caused by $\mathcal{T}^{\textrm{D}}_{\textrm{e}}$.
In order to discuss the permeability due to $\mathcal{T}^{\textrm{D}}_{\textrm{e}}$, we consider when the momentum has the $y$-component ($\bm{q}= q\bm{y}$), whose direction is parallel to the applied static magnetic field.
That means that the spin structure we consider is a longitudinal conical spin structure.
From the above equation, 
using $\langle M^{\parallel, x} \rangle_{\textrm{D}} = A_{q,\Omega} m_x - B_{q,\Omega} m_z$,
we obtain the magnetic permeability as  
\begin{align}
	\label{eq:5-3-69} %\ref{eq:5-3-69} 
	\biggl( \begin{matrix}  m_x \\ m_z \end{matrix} \biggr) 
	& = 
	\biggl( \begin{matrix}  \chi_{xx}^{\textrm{D}} & \chi_{xz}^{\textrm{D}} \\  \chi_{zx}^{\textrm{D}} & \chi^{\textrm{D}}_{zz}  \end{matrix} \biggr) 
	\biggl( \begin{matrix}  h_x \\ h_z \end{matrix} \biggr).
\end{align} 
Here the longitudinal and transverse magnetic permeability are given by 
\begin{align} 
	\label{eq:5-3-70} %\ref{eq:5-3-70}
\chi_{xx}^{\textrm{D}}(q, \Omega) & = 
	\frac{(\omega_H + i \alpha \Omega) \omega_M }
		{[\omega_H + i\alpha \Omega ] [ \omega_H + i (\alpha + \tilde{\kappa}_{q,\Omega}) \Omega] - \zeta_{q,\Omega} \Omega^2}, 
	\\ 
	\label{eq:5-3-71} %\ref{eq:5-3-71}
\chi_{zz}^{\textrm{D}}(q, \Omega)
	& = \frac{\omega_H + i (\alpha + \tilde{\kappa}_{q,\Omega}) \Omega }{\omega_H + i\alpha \Omega}\chi_{xx}^{\textrm{D}},
	\\
	\label{eq:5-3-72} %\ref{eq:5-3-72}
\chi_{xz}^{\textrm{D}}(q, \Omega) & =
	- \zeta_{q,\Omega} \chi_{zx}^{\textrm{D}}.
\end{align} 
The obtained permeability is different from that  in Eqs. (\ref{eq:5-2-59})-(\ref{eq:5-2-61}).
The difference is caused by coefficients $\tilde{\kappa}_{q,\Omega}$ and $\zeta_{q,\Omega}$: 
\begin{align} 
	\label{eq:5-3-73} %\ref{eq:5-3-73}
\tilde{\kappa}_{q,\Omega} & = \kappa ( 1- q^2 \ell^2 A_{q,\Omega}  ), 
	\\
	\label{eq:5-3-74} %\ref{eq:5-3-74}
\zeta_{q,\Omega} & = 1 +  \kappa q^2 \ell^2 B_{q,\Omega}.
\end{align} 
The $\tilde{\kappa}_{q,\Omega}$ and $\zeta_{q,\Omega}$ depend on $q\ell$ and $\Omega \tau$.
If $q=0$, one can demonstrate $\tilde{\kappa}_{q,\Omega} = \kappa$ and $\zeta_{q,\Omega}=1$, and 
the magnetic permeability in Eqs. (\ref{eq:5-3-70})-(\ref{eq:5-3-72}) equal to that in Eqs. (\ref{eq:5-2-59})-(\ref{eq:5-2-61}), respectively.
Figure 2(a) indicates the $\Omega \tau$ dependence of $\tilde{\kappa}/\kappa$ for several momentum $q\ell$.
The parameter $\tilde{\kappa}/\kappa$ approaches to 0 from $\tilde{\kappa}/\kappa=1$ with increasing $\Omega\tau$:
In the case for $\tilde{\kappa}/\kappa =0$, $\mathcal{T}_e^\textrm{L}$ and $\mathcal{T}_e^\textrm{D}$ are canceled out each other, and the spin torque $\mathcal{T}_e$ vanishes. 
On the other hand,  $\tilde{\kappa}/\kappa =1$ 
means that 
$\mathcal{T}_e^\textrm{D}$ is zero and $\mathcal{T}_e^\textrm{L}$ is nozero. 
The relation $\tilde{\kappa}/\kappa$ significantly changed when $(q\ell)^2 \sim \Omega \tau$ is satisfied.
Figure 2(b) shows the $q\ell$ dependence of $\zeta$ for several angular frequency of the applied ac magnetic field $\Omega$, where we take a realistic relaxation time ($\tau=0.087$ ps \ \cite{rf:Taskin12a}). 
The magnitude of $\zeta$ changes around $q\ell \sim \sqrt{\Omega \tau}$ 
and approaches to $\zeta =1$ with increasing $q\ell$.
%%%%%%%%%%%%%%%%
%%%%%%%%%%%%%% FIG2
\begin{figure}[tbp] \label{fig:2} %\ref{fig:2} 
\includegraphics[scale=.42]{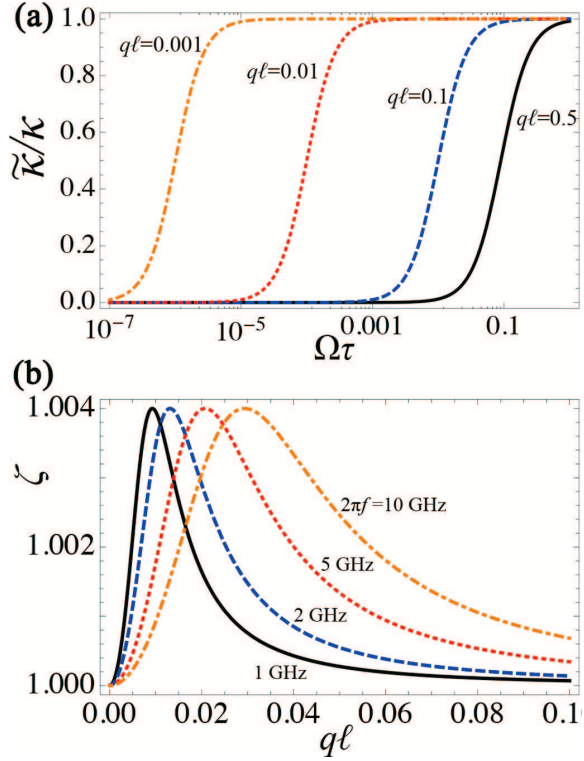} 
\caption{(Color online) 
(a) The $\Omega \tau$ dependence of $\tilde{\kappa}/\kappa$ for several 
 $q\ell$.
(b) The $q\ell$ dependence of the $\zeta$ for several frequency ($2\pi f=1,2, 5, 10$ GHz) in the fixed relaxation time ($\tau=0.087$ ps).
%: the solid lines correspond to $q\ell=0.1$, the dashed lines correspond to $q\ell=0.01$, and the dashed solid line correspond to $q\ell=0.03$.
%%%%
}
\end{figure}

%%%%%%%%%%%%%%%%
%     The figure 3.(a)-(b)     %
%%%%%%%%%%%%%%%%
Figure 3 (a) and (b) show the $\Omega \tau$ dependence of the real and imaginary part of the longitudinal magnetic permeability, $\textrm{Re} [\chi_{xx}^{\textrm{D}}]$ and $-\textrm{Im} [\chi_{xx}^{\textrm{D}}]$, respectively.
In Figs. 3(a) and (b), we choose realistic parameters of the TI: $\ell = 40$nm, $v_{\textrm{F}} = 4.6 \times 10^5$m/s\cite{rf:Taskin12a}, $\tau=0.087$ ps, $\xi = 0.003$, and $\tilde{v}_\textrm{F}/v_\textrm{F} = 0.997$.
Besides, we choose material parameters of the ferromagnets\cite{rf:Shiomi14}: $\alpha=0.015$, $J_{sd}\sim 6$meV, $S^\parallel \sim 0.3$, and Fermi wavenumber $k_\textrm{F}=3.9\times 10^{8} \textrm{m}^{-1}$. 
Then, $\kappa=0.008$ is obtained. 
The frequency of the magnetization $\omega_M/(2\pi) = f_M = 0.28$ GHz in the MI.
The magnitude of the  frequency is evaluated 
by the material parameters of the permalloy\cite{rf:Shiomi14}. 

We plot the $\textrm{Re} [\chi_{xx}^{\textrm{D}}]$ and $-\textrm{Im} [\chi_{xx}^{\textrm{D}}]$ functions as the frequency of the ac magnetic field for several momentum of the localized spin, when we take the frequency of the static magnetic field $\omega_H/(2\pi)=f_H = 1.6$ GHz.
The magnitude of $\Omega \tau$ can be estimated as $\Omega \tau \sim \omega_H \tau = 8.6 \times 10^{-3}$.
For $q\ell<0.03$, the permeability $\textrm{Re} [\chi_{xx}^{\textrm{D}}]$ and $-\textrm{Im} [\chi_{xx}^{\textrm{D}}]$ are not dramatically 
changed from that  without diffusion, $\textrm{Re} [\chi_{xx}]$ and $-\textrm{Im} [\chi_{xx}]$, respectively. 
The reason is due to the profile of $\tilde{\kappa}/\kappa$ and $\zeta$:
Both $\tilde{\kappa}/\kappa$ and $\zeta$ are about 1 within $q\ell<0.03$ in $\Omega \tau \sim 8.6 \times 10^{-3}$.
While $q\ell$ is near $q\ell =0.03$, $\chi_{xx}^\textrm{D}$ changes: 
The magnitude of $\chi_{xx}^\textrm{D}$ increases from that of $\chi_{xx}$. 
Besides,  the resonant frequency of $\chi_{xx}^\textrm{D}$ increases from that of $\chi_{xx}$ [see $q\ell =0.03$ in Figs. 3 (a)-(b)].
The change of the resonant frequency is also shown in Fig. 4 (a).
We find that when $(q\ell)^2 \sim \Omega \tau$ is satisfied, $\tilde{\kappa}/\kappa$ and $\zeta$ deviate from 1, as shown in Figs. 2(a)-(b).
After increasing $q\ell$ from $q\ell=0.03$, 
the shifted resonant frequency gets back again.
The magnitude of the permeability increases with increasing $q\ell$.
%%%%%%%%%%%%%%%%
%     The figure 3.(c)-(d)     %
%%%%%%%%%%%%%%%%
%Figure 3. (c)-(d) indicate the plot of $\chi_{xx}' $ and $\chi_{xx}'' $ in the relaxation time ($\tau=8.6$ ps)\cite{rf:Chen09}. Then, we chose $\kappa=0.8$ because of $\kappa \propto \tau$. 
%The profile of the $\chi_{xx}' $ and $\chi_{xx}'' $ are tend to be similar to that in $\tau=0.087$ ps:
%The magnitude of the permeability decreases with increasing $q\ell$ in whole $q\ell$.
%While the resonance frequency of the permeability decreases with increasing $q\ell$ within $0<q\ell<0.3$.
%With increasing $q\ell$ over $q\ell<0.3$, 
%the changed resonance frequency approaches to get back again.
%The change of its magnitude and resonance frequency dramatical changed when $q\ell $ is near 0.3 and $\Omega \tau \sim (q\ell)^2$ is satisfied.
%%%%%%%%%%%%%%%%
%%%%%%%%%%%%%% FIG3
\begin{figure}[tbp] \label{fig:3} %\ref{fig:3} 
\includegraphics[scale=.6]{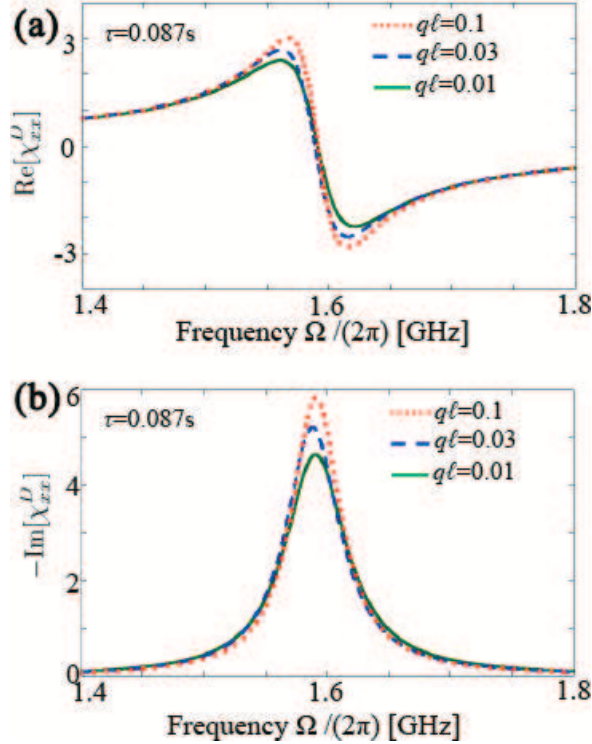} 
\caption{(Color online)  
(a)-(b) The frequency dependence of the real part and imaginary part of the longitudinal permeability, Re$[\chi_{xx}^\textrm{D}]$ and $-\textrm{Im}[\chi_{xx}^\textrm{D}]$, for a fixed Gilbert damping constant ($\alpha=0.015$), the relaxation time [$\tau=0.087$ps], the anisotropic damping constant [$\kappa=0.008$], and the frequency [$f_H=1.6$ GHz and $f_M=0.28$ GHz] for several momentum $q\ell$.
%(c) The $q\ell$ dependence of the resonance frequency and the half-width value.
%: the solid lines correspond to $q\ell=0.1$, the dashed lines correspond to $q\ell=0.01$, and the dashed solid line correspond to $q\ell=0.03$.
%%%%
%%%%%%
}
\end{figure}
%%%%%%%%%%%%
%     The figure 3 (c)    %
%%%%%%%%%%%%
\begin{figure}[tbp] \label{fig:4} %\ref{fig:4} 
\includegraphics[scale=.45]{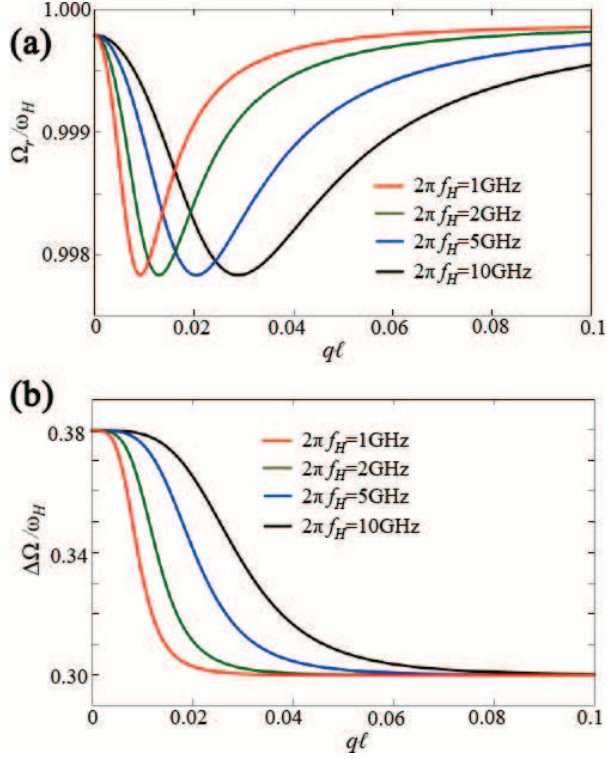} 
\caption{(Color online)  
The $q\ell$ dependence of the resonant frequency (a) and the half-width value (b) normalized by the angular frequency of the applied magnetic field for several frequencies ($2\pi f_H=$1, 2, 5, and 10 GHz) in $2\alpha+\kappa=0.038$ and $2\alpha=0.030$.}
\end{figure}

The half-width value is also changed by the magnitude of $\textrm{Im}[\chi_{xx}^\textrm{D}]$ for several momentum $q\ell$. 
This trade off between the half-width value and $q\ell$ for several frequency ($2\pi f_H = 1$, 2, 5, and 10 GHz ) is shown in Fig. 4(b). 
In the case for $\tau=0.087$ ps, the half-width value $\Delta \Omega$ significantly changed around the $q\ell$, which satisfies $q\ell \sim \sqrt{\Omega \tau}$ (e.g., $q\ell=0.03$ in $\omega_H = 10$ GHz).  
Figure 4(a) indicates $q\ell$ dependence of the resonant frequency rate $\Omega_r/{\omega_H}$.
The $\Omega_r/{\omega_H}$ changed from $\Omega_r/{\omega_H}\sim 1$ into $\Omega_r/{\omega_H}=0.998$ around the $q\ell$, which satisfies $q\ell \sim \sqrt{\Omega \tau}$.

We discuss the role of diffusive spin torque $\bm{\mathcal{T}}^{\textrm{D}}_{\textrm{e}}$ from Figs. 4(a) and (b).
Figure 4(a) shows that the resonant angular frequency $\Omega_r$ tends to decrease with increasing $q\ell$.
The decrease of $\Omega_r$ can be caused by the increase of $\zeta_{q,\Omega}$, and the change of $\zeta_{q,\Omega}$ is caused by  $\bm{\mathcal{T}}_{\textrm{e}}^{\textrm{D}}$.
Therefore, we expect that $\bm{\mathcal{T}}_{\textrm{e}}^{\textrm{D}}$ plays a role as a field-like torque to shift the resonant frequency. 
Then, Fig.4 (b) indicates that  
the half-width value tends to decrease with increasing  $\tilde{\kappa}_{q,\Omega}$ in Eq. (73), which is caused by $\bm{\mathcal{T}}^{\textrm{D}}_{\textrm{e}}$.
It means that the damping of the magnetization dynamics is reduced by $\bm{\mathcal{T}}_{\textrm{e}}^{\textrm{D}}$ on the disordered surface of the doped TI.
%%%%%%%%%%%%%%%%%%%%%%%%%%%%%%%%%%%%%%%%%%%%%%%%%%
%%%%%%%%%%%%%%%%%%%%%%%%%%%%%%%%%%%%%%%%%%%%%%%%%%%%%%%%%
Thus, $\bm{\mathcal{T}}^{\textrm{D}}_{\textrm{e}}$ behaves  
both the damping torque and  the field-like torque.
The role of the spin torque in several spin structures are summarized in the Table I.
From the table I, we can expect to distinguish the half-width value contributed from $\mathcal{T}_e^{\textrm{L}}$ and $\mathcal{T}_e^{\textrm{D}}$ by tuning of the magnitude of the applied magnetic field $\bm{H}_0$.
The reason is why the longitudinal spin structure changes an spatial uniform ferromagnetic structure  if when we apply a strong magnetic field, which broke the longitudinal spin structure. 
Then, $q\ell$ of the spatial uniform spin structure can be regarded as $q\ell =0$.
Therefore, we expect that if when we apply the strong applied magnetic field in the longitudinal spin structure, the half-width value with a finite $q\ell$ changes into the half-width value with $q\ell=0$ as $\Delta\Omega/\Omega_r=2\alpha+\tilde{\kappa}\to 2\alpha + \kappa$ [see Fig.4 (b)].

%%%%%%%%
% Table 1
%%%%%%%%
\begin{table}
\caption{A brief summary of the role of the spin torques  and the half-width value in several spin structures on the surface of the TI, when the direction of $\bm{H}_0$ and the propagation of $\bm{H}_{ac}$ are parallel to $y$-axis.}
\begin{center}
\begin{tabular}{lcccc}
\hline \hline
\ & Uniform  & Transverse conical ($q_y=0$)  & Longitudinal  conical ($q_y\neq0$)\\
%\multicolumn{1}{c}{Local driving field} & \multicolumn{1}{c}{Nonlocal driving field} \\ % ‹ÇŠ‹ì"®ê=local driving field "ñ‹ÇŠ‹ì"®ê
\hline
$\mathcal{T}_e^{\textrm{L}}$& Damping torque & Damping torque  & Damping torque \\
$\mathcal{T}_e^{\textrm{D}}$ & -- & -- & Damping torque and field-like torque 
\\
$\Delta \Omega/\Omega_r $ & $2\alpha +\kappa$ & $2\alpha +\kappa$ & $2\alpha +\tilde{\kappa}$ \\
\hline
\hline
\end{tabular}
\end{center}
\end{table}

%Magnitude of the spin density in the surface of the TI with the longitudinal spin structures. 
%Substituting $\bm{S}^\parallel = S \cos{(\bm{q}\cdot\bm{x}-\Omega t)}\hat{\bm{x}} +S_y \hat{\bm{y}} $ into Eq. (), we obtain the spin as 
%\begin{align}
%	\label{eq:5-3-74-3} %\ref{eq:5-3-74-3}
%\bm{s} & = 
%\end{align} 

%%%%%%%%%%%%%%%%%%%%%%%%%%%%%%%%%%%%%%%%%%%%%%%%%%%
%%%%%%%%%%%%%%%%%%%%%%%%%%%%%%%%%%%%%%%%%%%%%%%%%%%%%%%
%%%%%%%%%%%%%%
%% 	Contributed from an applied magnetic field   %%
%%%%%%%%%%%%%%

For magnetization dynamics due to the magnetic resonance, we need to apply magnetic fields in the MI/TI junction.
Then, the spin-charge generation and transport are triggered not only by the magnetization dynamics, but also by the applied magnetic field.
We will estimate when the contribution due to the magnetic field can be relevant. 
The contribution from the applied magnetic field can be described by the Zeeman effect, $\mathcal{H}_{\textrm{Z}} = - 2\hbar \gamma \int d\bm{x} \bm{B}\cdot \bm{s}$, 
where $\bm{B}$ is the applied magnetic field and couples with conduction electrons spin on the surface of the TI.
The contribution from $\bm{B}$ can be treated within the same formalism in sections 3 and 4 by replacing $\bm{S} \to \bm{S} + 2(\hbar \gamma/ J_{sd}) \bm{B}$ in Eq. (\ref{eq:2-0-3}).
As a result, the spin-charge generation and transport due to $\mathcal{H}_{\textrm{Z}}$ and $\mathcal{H}_{sd}$ are obtained by replacing $\bm{S} \to \bm{S} + 2(\hbar \gamma/ J_{sd}) \bm{B}$ in Eqs. (\ref{eq:3-2-27}), (\ref{eq:3-3-32}), (\ref{eq:4-1-39})-(\ref{eq:4-1-40}), and (\ref{eq:4-2-43})-(\ref{eq:4-2-44}).
We expect that the contribution from $\mathcal{H}_{\textrm{Z}}$ can be ignored compared with that from $\mathcal{H}_{sd}$, when the energy scale of the Zeeman effect is smaller than that of the exchange energy on the surface of the TI as $2 \hbar \gamma |\bm{B}| / (J_{sd} |\bm{S}|) \ll 1$. 
The $|\bm{B}|/ J_{sd}$ value can be estimated by $|\bm{B}| / J_{sd} \ll 1/(2 \hbar \gamma) \sim 4 \times 10^4$ T$\cdot$eV$^{-1}$ in $\bm{S} \sim 1$.
Then, the magnitude of the exchange coupling $J_{sd}$ can be estimated.
Using realistic parameters, the mean-free path $\ell = 40$nm, Fermi velocity $v_{\textrm{F}} = 4.6 \times 10^5$m/s \cite{rf:Taskin12a}, we obtain the $\hbar/\tau \sim 15$meV and $J_{sd} \ll 15$ meV, which is requested for the perturbation condition $J_{sd} \tau /\hbar \ll 1$. 
Then, if $|\bm{B}|\gg 10$T is satisfied, we need to consider the contribution from $\mathcal{H}_{\textrm{Z}}$.

%%%%%%%%%%%%%%
%% 	Discussion part  Order estimation %%
%%%%%%%%%%%%%%

%%%%%%%%%%%%%%
%% 	Discussion part 3  %%
%%%%%%%%%%%%%%

At the end of this section, we estimate the spin density due to the dynamics of the longitudinal spin structure $\bm{S} = S^\parallel (\sin\theta \cos\Omega t, 1, \sin\theta \sin\Omega t)$ in the case of $\theta\ll1$ around resonance angular frequency $\Omega \sim 1\times 10^{10}$s$^{-1}$.
The magnitude of the spin depends on the regime of $q^2 \ell^2 \ll \Omega \tau$ or $q^2 \ell^2 \gg \Omega \tau$.
In $q^2 \ell^2 \ll \Omega \tau$ regime, the magnitude of the nonlocal spin can be negligible small compared with that of the local term, and the spin is estimated by $|\bm{s}| \sim 1.6 \times 10^{-10}$\AA$^{-2}$ at $\theta\sim 0.1$ rad in the FMR.
Then, we find that the magnitude of the spin due to the spin-pumping is smaller than that of the spin due to the applied electric field\cite{rf:Misawa11}. 
On the other hand, in $q^2 \ell^2 \gg \Omega \tau$ regime, the local and nonlocal spin vanishes each other even in the presence of FMR.
From the results, we expect that the change of the magnitude of the spin dependent on $q^2 \ell^2 /\Omega \tau$ can be measurable for several applied magnetic fields, because the inhomogenous spin structure ($q\ell\neq 0$) changes into an uniform spin structure ($q\ell \sim 0$) by using an applied strong magnetic field.

\subsection{Spin current and charge current}
We will discuss the spin current on the surface of the disordered MI/TI junction compared with that in the FM/NM junction.
The spin current due to the spin-pumping $j_{i,\textrm{FM/NM}}^\alpha$ in the FM/NM junction is triggered by the magnetization dynamics as\cite{rf:Tserkovnyak02,rf:ohe07,rf:Takeuchi08} 
\begin{align}
	\label{eq:5-4-75} %\ref{eq:5-4-75}
j_{i,\textrm{FM/NM}}^\alpha & = b \nabla_i \partial_t S^\alpha + \mathcal{O}(S^2), %[ \nabla_i ( \bm{S} \times \partial_t \bm{S}) +   ( \bm{S} \times \nabla_i \partial_t \bm{S}) ]^\alpha
\end{align}
where $b$ is a coefficient dependent on materials.
It is similar to the spin current in Eq. (\ref{eq:5-4-75}), that the spin current is proportional to a time-dependent magnetization and the spin current vanishes when the magnetization is spatially uniform.
The direction of the spin $(\alpha)$ and the flow $(i)$ of the spin current in Eq. (\ref{eq:5-4-75}) are not related each other.
On the other hand, spin current on the surface of the TI, whose direction of spin and flow are perfectly perpendicular to each other.
The difference lies on the spin-orbit interaction.
The $j_{i}^\alpha$ in Eq. (\ref{eq:3-3-32}) includes the contribution 
of the spin-orbit interaction, which is absent 
in Eq. (\ref{eq:5-4-75}) does not.

Charge current due to the spin-pumping in the FM/NM junction is also given by the magnetization dynamics and Rashba type spin-orbit interactions. 
When the magnetization is spatially uniform, 
the charge current $\bm{j}_{\textrm{FM/NM}}$ becomes\cite{rf:ohe07,rf:Takeuchi08} 
\begin{align}
\label{eq:5-4-76} %\ref{eq:5-4-76}
\bm{j}_{\textrm{FM/NM}} & = \bm{\alpha} \times ( \bm{S} \times \partial_t \bm{S} ),
\end{align}
where $\bm{\alpha}$ is a constant vector including the contribution 
from the spin-orbit interaction. 
%and is parallel 
%to the direction of the inversion symmetry breaking in the system. 
The charge current in Eq. (\ref{eq:5-4-76}) is proportional to the second-order of the localized spin $\bm{S}$.
That is different from the charge current on the surface of TIs.
The charge current on the surface of the TI is proportional to the localized spin $\partial_t \bm{S}^\parallel$, as shown in Eq. (\ref{eq:4-2-44}).
The difference is due to the property of the localized spin:
the localized spin plays the role of the effective vector potential on the surface of the TI.
We note that the frequency dependence of these charge current 
is also different;
$\bm{j}_{\textrm{TI}} \propto \partial_t \bm{S}^\parallel$ oscillates with time of the localized spin in the FMR, 
but $\bm{j}_{\textrm{FM/NM}} \propto \bm{S}\times \partial_t \bm{S} $ does not.
For example, 
the ac current $\bm{j}_{\textrm{TI}} \propto (\cos{\Omega t}, \sin{\Omega t}, 0) $ is given when we apply the magnetic field parallel to the $z$ direction on the TI/MI junction.
On the other hand, the dc current $\bm{j}_{\textrm{FM/NM}} \propto \Omega (0,0,1)$ is obtained when we apply the magnetic field parallel to the $\bm{z}$ direction in the NM/FM junction.

%%%%%%%%%%%%%%
%%    	Summary           %%
%%%%%%%%%%%%%%

\section{Summary}
We have studied the spin-charge generation and transport due to the magnetization dynamics on the disordered surface of the doped TI/MI junction.
The spin current $j^\alpha_{s,i}$ is proportional to the charge density $\rho_e$ and the direction of its spin and its flow are perfectly perpendicular to each other, because of the spin-momentum locking on the surface of the TI. 
We have found that $j^\alpha_{i}$ and $\rho_e$ are induced by the time- and spatial-dependent of nonlocal magnetization dynamics, which is affected by nonmagnetic impurity scatterings on the disordered surface of the doped TI.
These results of $j^\alpha_{i}$ and $\rho_e$ are shown in Eqs. (\ref{eq:3-3-32}) and (\ref{eq:3-2-27}), respectively. 
$j^\alpha_{i}$ and $\rho_e$ is induced except when the magnetization dynamics is spatially uniform.
We have also shown the induced spin $\bm{s}$ and charge current density $\bm{j}$ due to the magnetization dynamics.
Because of the spin-momentum locking, the spin $\bm{s}$ and charge current density $\bm{j}$ are proportional to each other.
The $\bm{s}$ and $\bm{j}$ are generated not only by the local magnetization dynamics, but also by the nonlocal magnetization dynamics with the diffusion on the disordered surface of the doped TI.
These results of $\bm{s}$ and $\bm{j}$ are shown in Eqs. (\ref{eq:4-1-39})-(\ref{eq:4-1-40}) and (\ref{eq:4-2-43})-(\ref{eq:4-2-44}), respectively. 
A brief summary of the local and nonlocal $\rho_e$, $j_i^\alpha $, $\bm{s}$, and $\bm{j}$ due to the magnetization dynamics are represented in Table. II.
%
%
%
%
%%%%%%%%
%%%%%%%%
% Table 1
%%%%%%%%
\begin{table}
\caption{A brief summary of the charge, charge current, spin, and spin current density due to the magnetization dynamics on the disordered surface of the TI. 
These terms are driven by the effective electric field $\bm{\mathcal{E}}$.}
\begin{center}
\begin{tabular}{lcccc}
\hline \hline
\ & charge density $\rho_e$  & current density $j_i$  & spin density $s^\alpha$  & spin current density $j_i^\alpha$ \\
%\multicolumn{1}{c}{Local driving field} & \multicolumn{1}{c}{Nonlocal driving field} \\ % ‹ÇŠ‹ì"®ê=local driving field "ñ‹ÇŠ‹ì"®ê
\hline
Local term& -- & $\bm{z} \times \partial_t \bm{S}^\parallel$  & $\partial_t \bm{S}^\parallel$  & --\\
Nonlocal term & $[\bm{\nabla}\times \partial_t  \langle \bm{S}^\parallel \rangle_{\textrm{D}}]_z$ & $ \bm{\nabla} [\bm{\nabla}\times \partial_t  \langle \bm{S}^\parallel \rangle_{\textrm{D}}]_z $ & $(\bm{z}\times \bm{\nabla})[\bm{\nabla}\times \partial_t  \langle \bm{S}^\parallel \rangle_{\textrm{D}}]_z $ 
& $[\bm{\nabla}\times \partial_t  \langle \bm{S}^\parallel \rangle_{\textrm{D}}]_z$ \\
Driving field
& $\bm{\nabla}\cdot \langle \bm{\mathcal{E}} \rangle_{\textrm{D}}$ 
& $\bm{\mathcal{E}}$,  $\bm{\nabla}^2 \langle \bm{\mathcal{E}} \rangle_{\textrm{D}}$ 
& $\bm{z}\times \bm{\mathcal{E}}$,  $(\bm{z}\times  \bm{\nabla} ) [\bm{\nabla}\cdot \langle \bm{\mathcal{E}} \rangle_{\textrm{D}}]$ 
& $\bm{\nabla}\cdot \langle \bm{\mathcal{E}} \rangle_{\textrm{D}}$ \\
\hline
\hline
\end{tabular}
\end{center}
\end{table}
%%%%%%%%%%%
%
%
%
%
%
As a result, we have discussed the modification of the magnetization 
dynamics before and after these spin-charge generation and transport on the disordered surface of the doped TI. 
These spin-charge generation and transport can be detected from the half-width value of the magnetic permeability in the magnetic resonance in the MI/TI junction, as discussed in section V.
% added sentence 0129
The magnitude of a Gilbert damping constant $\alpha$ 
in ferromagnetic insulator is smaller than that in ferromagnetic metals.
Then, we can easily detect the change of the $f_H$ dependence of the resonant frequency and half-width value, which are shown in Figs. 4 (a)-(b). 
%and 
%Since the $f_H$ dependence of the resonance frequency and half-width value depend on the magnitude of $\alpha$, the change of the resonance frequency and half-width value, which are represented in Figs. 4 (a)-(b), can be detected from 
%
%
%Recently, high quality of the TI, such as Bi$_2$Se$_3$ and Bi$_2$Te$_3$, has been offered by developing technology\cite{rf:Taskin12a}.
%The magnetic doped TI has been suggested, such as the Mn-doped Bi$_2$Se$_3$ by the magnetic proximity effect through a deposited Fe overlayer\cite{rf:Vobornik}.
%%%%%%%%%%%%%%%%%%%%%%%%%%%%%%%%%%%%%%%%%%%%%%%%%%%%%%%%%%%%%%%%%%%%%%%%%%%%%%55%%%%%%%%%%%%%%%%%%%%%%%%%%%%%%%%%%%%%%%%%%%%%%%%%%%%%%%%%%%%%%%%%%%%%%%%%%5
%

The preparation of the hybrid system with  
the ferromagnetic insulator deposited on the surface of the TI, EuS/Bi$_2$Se$_3$, 
has been reported\cite{rf:Wei13}, where the magnetic 
moment of Eu locates at the interface between the EuS and Bi$_2$Se$_3$.
If the magnetization dynamics of the magnetic moment of Eu is triggered by an applied magnetic field,  
the spin density and charge current can be induced on the surface of the TI. 
Additionally, the magnetic distribution of Eu has a magnetic domain, which is spatially dependent on the position on the surface.
Therefore, when we move the magnetic domain by using an applied magnetic field, the charge density and the spin current are also triggered only around the magnetic domain. 
Recently, magnetic insulator with noncoplanar spin structure has been reported.  
%For example, an insulator magnet YIG has a spatially dependent magnetization dynamics in the presence of an applied magnetic field\cite{rf:Kajiwara}.
For example, magnetoelectric insulator Cu$_2$OSeO$_3$ has spatially dependent spin structure, and is called skyrmion, which is topologically protected magnetic spin vortex-like object\cite{rf:seki,rf:White}.
If one can prepare the vortex-like spin structure deposited on the surface of the TI and can trigger the magnetization dynamics of the skyrmion, 
we expect that the charge and spin currents are driven by the magnetization dynamics of the skyrmion.
Moreover, the spatial distributions of the charge density and the magnitude of the spin current depends on the position of the skyrmion,
because the spatial derivative of the localized spin depends on the positions in the skyrmion.
We comment that the induced spin and charge currents could be independent of the 
polarity of the skyrmion, since these current are triggered by the in-plane component of the localized spin [see Eqs. (\ref{eq:3-3-32}) and (\ref{eq:4-2-43})-(\ref{eq:4-2-44})]. 
Then, we expect that in the MIs deposited on the surface of the TI, the magnetization dynamics induces not only the local spin-charge generation and transport, but also the diffusive one. 
Our obtained results will enable the applications of TI nanomembrane in spintronics devices.

\section*{ACKNOWLEDGEMENTS}
%\acknowledgment
The authors would like to thank A. A. Golubov, A. Dutt, and A. Yamakaga for valuable discussions.
This work was supported by Grants-in-Aid for Young Scientists (B) (No. 22740222 and No. 23740236), by  Grants-in-Aid for Scientific Research on Innovative Areas ``Topological Quantum Phenomena'' 
(No. 22103005 and No. 25103709) from the Ministry of Education, 
Culture, Sports, Science, and Technology, Japan (MEXT), and by
the Core Research for Evolutional Science and Technology (CREST) of the Japan Science.
K.T. acknowledges support from a Grant-in-Aid for Japan Society for the Promotion of Science (JSPS) Fellows.

\appendix
\section{Derivation of Eqs. (\ref{eq:5-3-66})-(\ref{eq:5-3-67})} \label{sec:A}
We show details of calculation of coefficient $A_{q,\Omega}$ and $B_{q,\Omega}$ of nonlocal localized spin $\langle \bm{S} \rangle_{\textrm{D}}$.
The $\langle \bm{S} \rangle_{\textrm{D}}$ is given by 
\begin{align}	\label{eq:a-a-2} %\ref{eq:a-a-2}
\langle \bm{S} \rangle_{\textrm{D}} (\bm{x},t)
	& \equiv \iint dt' d\bm{x}' \mathcal{D} (\bm{x}-\bm{x}', t-t') \bm{S}(\bm{x}',t')
\end{align}
To substitute $\bm{S} =  S (1, 0, -i) e^{i(\bm{q}\cdot\bm{x} -\Omega t)} \propto e^{i(\bm{q}\cdot\bm{x} -\Omega t)}$ into the above equation, we calculate $\langle n  \rangle_{\textrm{D}} $:
\begin{align} \notag 
\langle n \rangle_{\textrm{D}} 
	& = \frac{1}{L^2}\iint dt'  d\bm{x}' \sum_{Q,\omega}
	\frac{e^{i[\omega (t-t')- \bm{Q}\cdot(\bm{x}-\bm{x}')]} e^{i(\bm{q}\cdot\bm{x'} -\Omega t')}}{Q^2\ell^2 + i\omega\tau}
	\\
	\label{eq:a-a-2} %\ref{eq:a-a-2}
	& = \frac{e^{i ( \bm{q}\cdot\bm{x} - \omega t)} }{q^2\ell^2 -i\Omega \tau}
\end{align}
The resulting $\langle S^x \rangle_{\textrm{D}} $ is obtained from the real part of $S\langle n \rangle_{\textrm{D}} $ in the above equation. 
In the same way, $\langle S^z \rangle_{\textrm{D}} $ is obtained from the real part of $-i S \langle n \rangle_{\textrm{D}} $.
Thus, from the above equation, the coefficient of $A_{q,\Omega}$ and $B_{q,\Omega}$ are derived as Eqs. (\ref{eq:5-3-66})-(\ref{eq:5-3-67}).

%%%%
\section{Charge conservation} \label{sec:Charge conservation} %\ref{sec:Charge conservation}
To check the validity of the spin current and the charge current we calculate, we use the charge conservation law
$\partial_t \rho_e + \bm{\nabla}\cdot\bm{j}=0$.
The charge density $\partial_t \rho_e$ is given by 
\begin{align} \label{eq:a-b-1} %\ref{eq:a-b-1}
\partial_t \rho_e 
	& =\frac{ e \nu_e J_{sd}\tau }{L^2} 
	\biggl[ \sum_{\bm{q},\Omega}e^{i[\Omega t-\bm{q}\cdot\bm{x}]}  
	 		\frac{ i \ell \Omega^2 } {q^2\ell^2 + i\Omega \tau} ( q_y S^x_{\bm{q},\Omega} -q_x S^y_{\bm{q},\Omega} ) \biggr], 
\end{align}
where $\ell= \tilde{v}_\textrm{F} \tau$ is the mean-free path.
%\begin{align} 
%\partial_t \rho_e
%	& = - \frac{2e \nu_e J_{sd} \tau}{ a} \sum_{\bm{q},\Omega}e^{i (\Omega t-\bm{q}\cdot\bm{x}) } 
%		\biggl[
%		\frac{ i \ell \Omega^2}{q^2 \ell^2 + i\Omega \tau}(q_y  S_{\bm{q},\Omega}^x-q_x  S_{\bm{q},\Omega}^y) 
%		\biggr].
%\end{align}
The resulting $\bm{\nabla}\cdot \bm{j}$ becomes 
\begin{align*}
\nabla_x j_x  &=  \frac{  e \tilde{v}_\textrm{F}   \nu_e J_{sd}\tau }{L^2}  \sum_{\bm{q},\Omega}e^{i[\Omega t-\bm{q}\cdot\bm{x}]} 
		\Omega \biggl[ \biggl\{ 1 -\frac{1}{2} \frac{q^2 \ell^2 }{q^2 \ell^2 + i\Omega \tau} \biggr\} q_x S^y_{\bm{q},\Omega}
	 + \frac{1}{2}\frac{ \ell^2  q^2  }{\ell^2 q^2 + i\Omega \tau}  q_y S^x_{\bm{q},\Omega} \biggr], 
	\\
\nabla_y j_y  &=  - \frac{  e \tilde{v}_\textrm{F}   \nu_e J_{sd}\tau }{ L^2}  \sum_{\bm{q},\Omega}e^{i[\Omega t-\bm{q}\cdot\bm{x}]} 
		\Omega \biggl[ \biggl\{ 1 -\frac{1}{2} \frac{q^2 \ell^2 }{q^2 \ell^2 + i\Omega \tau} \biggr\} q_y S^x_{\bm{q},\Omega}
	 + \frac{1}{2}\frac{ \ell^2  q^2 }{\ell^2 q^2 + i\Omega \tau}  q_x S^y_{\bm{q},\Omega} \biggr],
\end{align*}
and 
\begin{align}\notag
\nabla_x j_x + \nabla_y j_y   &=  \frac{ e \tilde{v}_\textrm{F}   \nu_e J_{sd} \tau}{ L^2}  \sum_{\bm{q},\Omega}e^{i[\Omega t-\bm{q}\cdot\bm{x}]} 
		\Omega \biggl[ \biggl\{ 1 - \frac{q^2 \ell^2 }{q^2 \ell^2 + i\Omega \tau} \biggr\} ( q_x S^y_{\bm{q},\Omega} - q_y S^x_{\bm{q},\Omega})
			\biggr].
		\\
		\label{eq:a-b-2} %\ref{eq:a-b-2}
		&=  \frac{ e  \nu_e J_{sd} \tau }{L^2}  \sum_{\bm{q},\Omega}e^{i[\Omega t-\bm{q}\cdot\bm{x}]} \biggl[ \frac{ i\Omega^2 \tilde{v}_\textrm{F}   \tau }{q^2 \ell^2 + i\Omega \tau}  ( q_x S^y_{\bm{q},\Omega} - q_y S^x_{\bm{q},\Omega})  \biggr] = - \partial_t \rho_e.
\end{align}
Therefore, the $\rho_e$ and $\bm{j}$ satisfy $\partial_t \rho_e + \bm{\nabla}\cdot\bm{j}=0$.


\begin{references}
\bibitem{rf:estimation}Using $\frac{\hbar}{2\tau}= \frac{1}{2}\pi\nu_e n_i u_0^2$, $\nu_e = \frac{\epsilon_\textrm{F}}{2\pi \hbar^2 \tilde{v}^2_\textrm{F}}$, and $\tilde{v}_\textrm{F}/v_\textrm{F}=(1+\xi)^{-1}$, 
we can obtain the relation: $\xi (1+\xi)^2 = \frac{\hbar}{4\pi \epsilon_\textrm{F}\tau}$.
Because of $\frac{\hbar}{\epsilon_\textrm{F}\tau}\ll1$, $\xi < \frac{\hbar}{4\pi \epsilon_\textrm{F}\tau}$ becomes negligible small as $\tilde{v}_\textrm{F}/v_\textrm{F} = 1 + o(\frac{\hbar}{\epsilon_\textrm{F}\tau})$.
Streakily speaking, there is no problem to take $\tilde{v}_\textrm{F} \simeq v_\textrm{F}$.
\bibitem{rf:discussion1}
If we do not take the replacement $v_{\rm{F}} \to \tilde{v}_{\rm{F}} $ in Eq. (\ref{eq:2-0-2}), the velocity operator on the surface of the TI becomes $\hat{v} = v_{\textrm{F}} (\bm{z}\times\hat{\bm{\sigma}})$.
Then, the charge current is given by the multiple of $v_\textrm{F}/\tilde{v}_{\rm{F}}$ in Eqs. (\ref{eq:4-2-43})-(\ref{eq:4-2-44}). 
The charge current does not satisfy the charge conservation law, which is shown in Appendix \ref{sec:Charge conservation}.
\bibitem{ref:matrix} $\hat{\Lambda}^{\textrm{ra}} = [1-\tilde{\Gamma}^{\textrm{ra}}]^{-1}$ is given when $max\{ \sum_{\nu} [\tilde{\Gamma}^{\textrm{ra}}]_{\mu \nu} \} <1$ is satisfied.
\end{references}
\end{document}